\documentclass[a4paper,11pt]{article}
\usepackage[utf8]{inputenc}
\usepackage{graphicx}
\usepackage{amsmath}
\pdfoutput=1
\usepackage{slashed,mathtools}
\usepackage{jcappub}
\usepackage{subfig}
\usepackage{tikz}
\usepackage{multirow}
\allowdisplaybreaks

\usepackage{physics}
\usepackage{hepnames}
\newcommand{\dbd}[2]{\ifmmode \frac{\textrm{d}#1}{\textrm{d}#2}\else $\textrm{d}#1/\textrm{d}#2$\fi}
\newcommand{\pbp}[2]{\ifmmode \frac{\partial#1}{\partial#2}\else $\partial#1/\partial#2$\fi}

\DeclareMathAlphabet{\mathpzc}{OT1}{pzc}{m}{it}
 \newcommand{\eV}{\text{e\kern-0.15ex V}\xspace}

 \newcommand{\GeV}{\text{G\eV}\xspace}
 \newcommand{\TeV}{\text{T\kern-0.1ex \eV}\xspace}

\usepackage{tikz,xcolor,hyperref}

\definecolor{lime}{HTML}{A6CE39}
\DeclareRobustCommand{\orcidicon}{\hspace{-1mm}
	\begin{tikzpicture}
	\draw[lime, fill=lime] (0,0) 
	circle [radius=0.16] 
	node[white] {{\fontfamily{qag}\selectfont \tiny \,ID}};
	\draw[white, fill=white] (-0.0525,0.095) 
	circle [radius=0.007];
	\end{tikzpicture}
	\hspace{-3mm}
}

\foreach \x in {A, ..., Z}{\expandafter\xdef\csname orcid\x\endcsname{\noexpand\href{https://orcid.org/\csname orcidauthor\x\endcsname}
			{\noexpand\orcidicon}}
}

\title{LHAASO Galactic Plane $\gamma$-rays Strongly Constrain Heavy Dark Matter}

\author[a,b]{Celine Boehm}
\author[c]{Ranjan Laha}
\author[d,e,a]{Tarak Nath Maity}
\affiliation[a]{School of Physics, The University of Sydney and ARC Centre of Excellence for Dark Matter Particle Physics, NSW 2006, Australia}
\affiliation[b]{The University of Edinburgh, School of Physics and Astronomy, EH9 3FD Edinburgh, UK}
\affiliation[c]{Centre for High Energy Physics, Indian Institute of Science, C.\,V.\,Raman Avenue, Bengaluru 560012, India}
\affiliation[d]{Theory Division, Saha Institute of Nuclear Physics, 1/AF, Bidhannagar, Kolkata-700064, India}
\affiliation[e]{Homi Bhabha National Institute, Training School Complex, Anushaktinagar, Mumbai-400094, India}
\emailAdd{celine.boehm@sydney.edu.au}
\emailAdd{ranjanlaha@iisc.ac.in}
\emailAdd{tarak.maity.physics@gmail.com}

\abstract{LHAASO, a ground-based observatory, is unveiling  new frontiers in our understanding of high-energy $\gamma$-rays and cosmic rays. It has recently observed high energy diffuse $\gamma$-rays from the Galactic plane in the TeV–PeV range. For the first time, we analyze this data to search for signatures of heavy decaying and annihilating dark matter in the mass range $10^5-10^{11}\,$GeV. We compute the expected photon flux from both Galactic and extragalactic dark matter, incorporating attenuation due to photon pair production. For the Galactic contribution, we include both prompt photons and secondary photons produced via inverse Compton scattering, accounting for electron/positron propagation. For the extragalactic component, in addition to the prompt and inverse Compton contributions, we also include cascade photons arising from inverse Compton scattering of pair-produced electrons and positrons. By combining all these contributions, we derive constraints on the dark matter parameter space. Our bounds for various two body Standard Model final states are strongest to date. This underscore LHAASO's capability to discover the nature of heavy dark matter.}

\begin{document}
\maketitle
\flushbottom

%%%%%%%%%%%%%%%%%
\section{Introduction}
\label{sec:intro}
%%%%%%%%%%%%%%%%
An invisible form of matter, known as dark matter (DM), is required to consistently explain a wide range of astrophysical and cosmological observations \cite{Cirelli:2024ssz}. It accounts for approximately $\sim 85\%$ of the total matter content of the Universe. While the existence of DM is well established, its fundamental nature remains unknown, posing one of the most significant challenges in modern science. The only property of DM that we are certain of is that it possess a mass. Under the simplest assumption that a single species constitutes an observationally relevant component of DM, the viable mass range spans from $\sim 10^{-21}\,$eV \cite{Zimmermann:2024xvd} up to $\sim 10^5$ solar masses\,\cite{Brandt:2016aco}. In this work, we focus on the search for heavy DM in the mass range $10^5 -10^{11}\,$GeV.

Probing such heavy DM is challenging for current collider-based experiments \cite{Boveia:2022jox}, as its production is kinematically suppressed at the currently accessible energies. Direct detection experiments, which typically search for DM scattering off  the Standard Model (SM) particles, also face limitations in this mass regime due to the significantly lower number density of heavy DM particles \cite{Carney:2022gse}. This leads to a negligible event rate, particularly for small DM–SM interaction cross sections. Indirect detection offers a promising complementary approach to test these candidates, wherein one searches for stable SM by-products resulting from DM annihilation or decay \cite{Slatyer:2021qgc} at energies where there are measurements and fluxes are computed using known processes. Space- and ground-based observatories can leverage these signals to discover otherwise inaccessible mass scales \cite{Murase:2012df, Murase:2012xs, Murase:2015gea, Kuznetsov:2016fjt, Hiroshima:2017hmy, Bhattacharya:2017jaw, Chianese:2017nwe, IceCube:2018tkk, Kachelriess:2018rty, Chianese:2018ijk, Arguelles:2019ouk, Ng:2020ghe, Chianese:2021htv, Esmaili:2021yaw, Maity:2021umk, Guepin:2021ljb, Skrzypek:2022hpy, Deliyergiyev:2022bvp, Montanari:2022buj, Das:2023wtk, Zhu:2023bex, Guo:2023axz, Fiorillo:2023clw, Song:2023xdk, Song:2024vdc, Munbodh:2024ast, Das:2024bed, Sarmah:2024ffy, IceCube:2023ies, Borah:2025igh, Khan:2025gxs, Murase:2025uwv, Barman:2025hoz, Jho:2025gaf, Kohri:2025bsn}. Among these byproducts, photons and neutrinos are particularly valuable messengers due to their lack of electric charge, which allows them to traverse cosmic distances without significant deflection. In this work, we use the latest Galactic plane high-energy diffuse $\gamma$-ray data-set from the Large High Altitude Air Shower Observatory (LHAASO) \cite{LHAASO:2024lnz} to search for potential signatures of heavy DM.

LHAASO is a ground-based observatory located at an altitude of approximately 4410 meters above sea level on the Haizi Mountain in Daocheng, Sichuan Province, China \cite{LHAASO:2019qtb}. It is designed to observe both high-energy cosmic rays (CRs) and  $\gamma$-rays by detecting the extensive air showers produced by their interactions in the atmosphere \cite{LHAASO:2023gne, LHAASO:2023rpg, LHAASO:2024knt, LHAASO:2025byy}. What makes LHAASO a uniquely powerful facility is its integration of multiple detection techniques:\,it employs scintillator arrays to detect the electromagnetic component of air showers, as well as Cherenkov radiation detectors using both water and atmosphere as the detection medium. These complementary approaches enable precise measurements of high-energy CRs and $\gamma$-rays across a broad energy range. For the detection of diffuse high-energy $\gamma$-rays, LHAASO utilizes two major detectors:\,(a) the Square Kilometer Array (KM2A) and (b) the Water Cherenkov Detector Array (WCDA). A key challenge in this search is the overwhelming background from CRs, which are approximately 3–4 orders of magnitude more abundant than the signal. KM2A addresses this challenge by exploiting the higher muon content typically associated with CR-induced air showers, enabling efficient discrimination against $\gamma$-ray events \cite{LHAASO:2023gne}. In contrast, WCDA distinguishes between the broader lateral shower profile of CRs and the more compact profile of $\gamma$-ray-induced showers \cite{LHAASO:2024lnz}. The combined measurements from KM2A and WCDA have provided unprecedented data on the diffuse $\gamma$-ray flux in the TeV to PeV energy range \cite{LHAASO:2024lnz}. Notably, in the TeV regime, LHAASO is the first experiment to report diffuse photon observations, effectively bridging the observational gap between Fermi Large Area Telescope (Fermi-LAT) at lower energies \cite{2012ApJ...750....3A} and Tibet AS$_\gamma$ at higher energies \cite{TibetASgamma:2021tpz}.

The origin of the observed diffuse $\gamma$-ray flux and its multimessenger implications are currently advancing our understanding of Galactic CRs \cite{Kato:2025gva, DeLaTorreLuque:2025zsv, Prevotat:2025ktr, Castro:2025wgf}. Given the well-established presence of CRs \cite{2018ApJ...854L...2M, 2015PhRvL.114q1103A, Alemanno:2021gpb, Yoon:2017qjx, CALET:2019bmh, CALET:2023nif, Atkin:2018wsp, 2024PhRvL.132e1002V, 2012APh....36..183A, Apel:2013uni, IceCube:2019hmk}, it is widely accepted that a significant fraction of the observed $\gamma$-ray flux arises from CR interactions with the interstellar medium (ISM) \cite{DeLaTorreLuque:2025zsv, Vecchiotti:2024kkz, Yan:2023hpt}. Although the LHAASO collaboration has argued for the presence of an additional component to fully account for their measured flux \cite{LHAASO:2024lnz}, subsequent theoretical studies have shown that CR interactions in the ISM alone may be sufficient to explain the LHAASO observations, while remaining consistent with diffuse $\gamma$-ray measurements from Fermi-LAT \cite{DeLaTorreLuque:2025zsv}. It is important to note that theoretical predictions of this CR-induced $\gamma$-ray flux are subject to several uncertainties. These include discrepancies in the measured CR proton spectra between IceCube \& IceTop \cite{IceCube:2019hmk} v/s KASCADE \cite{KASCADE:2005ynk} and KASCADE-Grande \cite{Apel:2013uni}, as well as variations in the assumed ISM parameters \cite{Vecchiotti:2024kkz}, CR cross-section, etc.,\,\cite{Vecchiotti:2024kkz}. Given the presence of this guaranteed but model-dependent CR induced high-energy background photons, this paper explores for first time the extent to which the latest LHAASO-WCDA Galactic plane results can be used to reveal or constrain possible contributions from DM.

We consider both the self-annihilation and decay of DM into two-body SM final states. For each of these scenarios, the resulting $\gamma$-ray signal receives contributions from both Galactic and extragalactic (EG) components \cite{Cirelli:2010xx}. In both cases, the total flux comprises of prompt photons, which are directly produced from DM annihilation or decay into SM particles, and secondary photons. Secondary photons arises from inverse Compton (IC) scattering, where electrons and positrons generated in DM interactions up-scatter low-energy background photons, including the cosmic microwave background (CMB), extragalactic background light (EBL), starlight (SL), and infrared (IR) photons. For the galactic IC photon flux, we incorporate the effects of electron spatial propagation, often neglected in earlier analyses \cite{Esmaili:2015xpa, Chianese:2019kyl, Leung:2023gwp, LHAASO:2022yxw}, by employing the formalism of Ref.\,\cite{Delahaye:2007fr}. We find that, in the case of heavy DM searches using LHAASO, electron propagation effects can be safely ignored due to the subdominant contribution of secondary IC photons. Additionally, for the EG DM-induced photon flux, we include contributions from electromagnetic cascades, which arise from the iterative interactions of pair-produced $e^\pm$ with the CMB and EBL via IC scattering. By accounting for all these components, we analyze the latest LHAASO data to search for both decaying and annihilating heavy DM. In the absence of a DM signal, we find that the resulting bounds can be up to a few times stronger than all previous constraints, thereby establishing promising avenues for probing heavy DM with high-energy $\gamma$-ray telescopes.  

The paper is organised as follows: in section \ref{sec:DM} we calculate galactic and EG high energy $\gamma$-ray flux from DM. In section \ref{sec:results} we present our results before concluding in section \ref{sec:conclusion}. 

%%%%%%%%%%%%%%%%%
\section{Photons from DM}
\label{sec:DM}
%%%%%%%%%%%%%%%%
DM annihilation or decay into various SM final states eventually produce photons. For simplicity, in this work we focus exclusively on the two-body decay and annihilation channels with 100\% branching ratios in each case, without assuming any specific particle physics model (see Refs.\,\cite{Feldstein:2013kka, Harigaya:2013pla, Harigaya:2014waa, Bhattacharya:2014yha, Rott:2014kfa, Dudas:2014bca, Daikoku:2015vsa, Kopp:2015bfa, Aisati:2015vma, Anchordoqui:2015lqa, Boucenna:2015tra, Ko:2015nma, Aisati:2015ova, Dev:2016qbd, Roland:2015yoa, Dev:2016uxj, ReFiorentin:2016rzn, DiBari:2016guw, Chianese:2016smc, Cata:2016epa, Bhattacharya:2016tma, Hiroshima:2017hmy, Borah:2017xgm, Chakravarty:2017hcy, Chianese:2017nwe, Dhuria:2017ihq, Sui:2018bbh, Lambiase:2018yql, Xu:2018qnd, Chianese:2018ijk, Kim:2019udq, Jaeckel:2020oet, Garcia:2020hyo, Azatov:2021ifm, Jaeckel:2021gah, An:2022toi, Allahverdi:2023nov} for the same). Galactic and EG DM annihilation/decay can produce photons from all SM final states. In the following, we discuss each of these contributions in detail.
%%%%%%%%%%
\subsection{Galactic photons from DM}
\label{subsec:galactic}
%%%%%%%%%
Galactic DM  annihilation or decay produces photons directly from the resulting SM states, commonly referred to as prompt photons. Additionally, $e^\pm$ produced in these interactions can up-scatter background photons, such as those from the CMB, EBL, SL and IR radiation, via IC scattering\footnote{Synchrotron and bremsstrahlung photons produced from $e^\pm$ will contribute sub-dominantly for our DM mass range \cite{Munbodh:2024ast}.}, generating secondary photons with somewhat lower energies than the prompt ones. These processes have been discussed in detail in Ref.\,\cite{Cirelli:2010xx}. For completeness, we summarize them below.
%%%%%%%
\subsubsection{Prompt photons}
\label{subsubsec:GalPrompt}
%%%%%%
The differential prompt photons flux for DM of mass $m_\chi$ is
\begin{equation}
 \frac{\mathrm{d}^2 \phi_\gamma}{\mathrm{d} E_\gamma \, \mathrm{d} \Omega}\bigg\rvert_{\rm G}^{\rm p} = \frac{r_\odot}{4\pi \eta}
  \left(\frac{\rho_\odot}{m_\chi}\right)^\eta \mathcal{P}_\eta^\gamma (E_\gamma) \int \frac{\dd \Omega}{\Delta \Omega}\frac{\dd s}{r_\odot} \left(\frac{\rho_{\chi}(r(s,b,\ell))}{\rho_\odot}\right)^\eta \, e^{-\tau_{\gamma \gamma}^{\rm G}(E_\gamma,s,b,\ell)} \mathcal{B}^{\rm G}_\eta \,,
\label{eq:promptG}
\end{equation}
where $\eta = 1$ for decay and $\eta = 2$ for self-conjugate DM annihilation. The distance from the Milky Way (MW) center to the Sun is $r_\odot = 8.3\,\text{kpc}$, and the local DM density is taken to be $\rho_\odot = 0.3\,\GeV/{\rm cm}^3$. Line-of-sight distance and the angular region of observation are represented by $s$ and $\Omega$, respectively. The position-dependent DM density of the MW halo is expressed by $\rho_\chi (r)$. The total solid angle corresponding to the relevant observation region is denoted by $\Delta\Omega$. The factor $\mathcal{P}_\eta^\gamma$ is:
\begin{equation}
  \mathcal{P}_1^\gamma (E_\gamma) = \frac{1}{\tau_\chi} \dv{N_\gamma}{E_\gamma} ~ \text{(decay)} \quad \quad
  \mathcal{P}_2^\gamma (E_\gamma) = \langle \sigma v \rangle \dv{N_\gamma}{E_\gamma} ~ \text{(annihilation)},
\label{eq:p_etaDM}  
\end{equation}
with $\dd{N_\gamma}/\dd{E_\gamma}$ denoting the photon spectrum, adapted from \texttt{HDMSpectra}\,\cite{Bauer:2020jay} which includes electroweak corrections. The DM lifetime and the velocity averaged annihilation cross section are represented by $\tau_{\chi}$ and $\langle \sigma v \rangle$, respectively. The attenuation of the photon flux due to pair production processes is accounted by $e^{-\tau_{\gamma \gamma}^{\rm G}}$, where $\tau_{\gamma \gamma}^{\rm G}$ is the Galactic optical depth \cite{Gould:1967zzb}. The Galactic boost factor is denoted by $\mathcal{B}^{\rm G}_\eta$; for decaying DM, $\mathcal{B}^{\rm G}_1 = 1$, while for annihilating DM, the boost factor $\mathcal{B}^{\rm G}_2$ depends on DM halo properties. Detailed information about all these factors is provided below.

We choose Navarro–Frenk–White (NFW) profile \cite{Navarro:1995iw,Navarro:1996gj} for the DM density\footnote{Since LHAASO is looking away from Galactic center, other choices of the DM profile \cite{Haud:1986yj,Burkert:1995yz} change our results minimally.}
\begin{equation}
\rho_\chi (r) \,=\, \rho_\odot \, \left(\dfrac{r_\odot}{r} \right) \, \left(\dfrac{1 \,+\, (r_\odot/r_s)}{1 \,+\, (r/r_s)} \right)^2,
\label{eq:NFW}
\end{equation}
with the scale radius $r_s = 20\,$kpc. The galactrocentric radius, $r=\sqrt{s^2+r_\odot^2 - 2 s r_\odot \cos{b} \cos{\ell}}$, where $b$ and $\ell$ are galactic latitude and longitude, respectively.

High-energy photons ($\gamma$) produced from DM interactions can possess sufficient energy to interact with background photons ($\gamma_{i}$), leading to electron–positron pair production via the process $\gamma + \gamma_i \to e^+ e^-$. The relevant background photon fields include the CMB, EBL, SL, and IR radiation. The CMB consists of relic photons from the epoch of last scattering. The EBL originates from the integrated light emitted by all stars in the observable universe, with a portion of this light absorbed and re-emitted by interstellar dust at longer wavelengths. While various models exist to describe the EBL, in this work we adopt the observationally motivated model of Saldana-Lopez et al \cite{Saldana-Lopez:2020qzx} (see Refs.\,\cite{2011MNRAS.410.2556D, Franceschini:2017iwq, 2022ApJ...941...33F} for other EBL models). The SL and IR components refer to the radiation emitted by Galactic stars and the corresponding infrared re-emission from interstellar dust. These two components exhibit non-trivial spatial distributions due to the structured distribution of stars and dust in the MW. In this work, we adopt the model presented in Ref.\,\cite{Vernetto:2016alq}, supplemented by the relevant molecular transition lines described in Ref.\,\cite{Weingartner:2000cc} to describe the SL and IR photon fields (see Ref.\,\cite{2011CoPhC.182.1156V, Porter:2005qx} for corresponding {\tt Galprop} based model).

The Galactic optical depth of high energy photons can be evaluated using:
\begin{equation}
    \tau_{\gamma \gamma}^{\rm G}(E_\gamma, s,b,\ell) = \sum_i \int_0^s \dd s^\prime \int \dd \epsilon \,  \dd\theta \, \sigma_{\gamma \gamma} (E_\gamma,\epsilon) n_i(\epsilon,s^\prime, b,\ell) \frac{1-\cos{\theta}}{2} \sin{\theta}.
   \label{eq:tau_G} 
\end{equation}
Here, the differential background photon density of energy $\epsilon$ is denoted by $n_i$ which we sum over for all the relevant background photons, i.e., CMB, EBL, SL and IR. The angle between the momentum of the high-energy photon and the background photon is represented by $\theta$. The densities of the CMB and EBL are isotropic, i.e., independent of $s^\prime$, $b$, and $\ell$, which allows for a further simplification of Eq.\,\eqref{eq:tau_G} \cite{Esmaili:2015xpa}. The pair production cross section is\,\cite{Breit:1934zz}
\begin{equation}
    \sigma_{\gamma \gamma} = \frac{3 \sigma_T}{16} \left(1-\beta^2\right)\left[(3-\beta^4) \ln \left(\frac{1+\beta}{1-\beta}\right) -2 \beta (2-\beta^2)\right],
\end{equation}
where the Thomson scattering cross section is given by $\sigma_T = 8\pi \alpha^2/m_e^2$, with electron mass $m_e$ and the fine-structure constant $\alpha$. The dimensionless variable $\beta$ is defined as $\beta = \sqrt{1 - 1/\tilde{s}}$, where $\tilde{s} = s_0 (1 - \cos{\theta})/2$ and $s_0 = \epsilon E_\gamma / m_e^2$, characterizing the center-of-mass energy of the $\gamma\gamma_i$ system.

Within the framework of hierarchical structure formation, the DM distribution is expected to exhibit clumpy features due to the presence of smaller substructures. Since the annihilation rate scales with the square of the DM density, i.e., $\langle \rho_\chi^2 \rangle$, the presence of these substructures can enhance the overall annihilation rate because $\langle \rho_\chi^2 \rangle > \langle \rho_\chi \rangle^2$.  This enhancement is quantified by the so-called boost factor. In contrast, for decaying DM, the flux scales linearly with $\langle \rho_\chi \rangle$, and thus no boost factor is present. For the Galactic case the boost factor is evaluated using\,\cite{Kamionkowski:2010mi}
\begin{equation}
  \mathcal{B}^{\rm G}_2 = f_{\rm sm}  e^{\delta_f^2} + (1-f_{\rm sm})\frac{1-\alpha_K}{1+\alpha_K}\left[\left(\frac{\rho_{\rm max}}{\rho_\chi(r)}\right)^{1-\alpha_K}-1\right],
  \label{eq:boostG}
\end{equation}
where the fraction of smooth component of the DM halo is $f_{\rm sm} = 1-\kappa \left(\rho_\chi(r)/\rho_\chi(100\,{\rm kpc}) \right)^{-0.26}$. Calibrating Eq.\,\eqref{eq:boostG} with results from numerical $N$-body simulations, Ref.\,\cite{Kamionkowski:2010mi} determines the parameters as $\delta_f=0.2,~\alpha_K = 0,~ \kappa=0.007$. For the LHAASO dataset, the inclusion of this boost factor results in an $\mathcal{O}(1)$ enhancement in the predicted annihilation flux.
%%%%%%%
\subsubsection{Inverse Compton photons}
\label{subsubsec:GalIC}
%%%%%%
Apart from the prompt photon flux discussed in the previous section, $e^\pm$ produced from DM interactions can up-scatter off background photons to generate secondary $\gamma$-rays. This contribution becomes particularly relevant at lower energies, where the flux of prompt photons tends to be subdominant. For a given dataset, the inclusion of the IC contribution effectively improves the sensitivity for higher DM masses compared to prompt photon signals alone\,\cite{Cirelli:2009vg,Meade:2009iu, Cohen:2016uyg,Cirelli:2020bpc, Cirelli:2023tnx}.

The differential IC photon flux from DM is given by \cite{Cirelli:2009vg,Meade:2009iu}
\begin{equation}
\frac{\mathrm{d}^2 \phi_\gamma}{\mathrm{d} E_\gamma \, \mathrm{d} \Omega}\bigg\rvert_{\rm G}^{\rm IC} = \frac{r_\odot}{4\pi E_\gamma^2 \eta} \left(\frac{\rho_\odot}{m_\chi}\right)^\eta \int \dd E_e^s \, \mathcal{P}_\eta^e (E_e^s) \, \int \frac{\dd \Omega}{\Delta \Omega} I^{\rm IC}_\eta(E_\gamma,E_e^s,b,\ell),
\label{eq:galIC}
\end{equation}
where $\mathcal{P}_\eta^e$ can be followed from Eq.\,\eqref{eq:p_etaDM} with photon spectrum replaced by the injected source electron spectrum with energy $E_e^s$.  The IC halo function is given by \cite{Lacroix:2013qka, Lacroix:2014eea, Cirelli:2010xx}
\begin{equation}
I^{\rm IC}_\eta(E_\gamma,E_e^s,b,\ell) = 2 E_\gamma \int \frac{\dd s}{r_\odot} \mathcal{B}^{\rm G}_\eta  \int_{E_\gamma}^{E_e^s} \dd E_e \frac{P_{\rm IC}^{\rm G}(E_\gamma,E_e,s,b,\ell)}{b^{\rm G}(E_e,s,b,\ell)} \tilde{I}_\eta(E_e^s,E_e,s,b,\ell) \, e^{-\tau_{\gamma \gamma}^{\rm G}}.
\label{eq:I_IC}
\end{equation}
where  $P_{\rm IC}^{\rm G}$ is the power radiated by an electron of energy $E_e$ into a photon of energy $E_\gamma$ at any point in the Galaxy. The total energy loss that electrons experience is denoted by $b^{\rm G}$ and dominated by IC, synchrotron. The $e^\pm$ halo function, $\tilde{I}_\eta(E_e^s, E_e, s, b, \ell)$, accounts for propagation effects and quantifies the contribution of an electron being observed with energy $E_e$ at a given location in the Galaxy after being injected with energy $E_e^s$ from all other points.

The IC power is given by
\begin{equation}
P_{\rm IC}^{\rm G}(E_{\gamma},E_e,s,b,\ell) =  \sum_i \frac{3 \sigma_{T} E_{\gamma} }{4 \gamma^2} \int^1_{\frac{1}{4\gamma^2}} \dd{q} \left[1-\frac{1}{4q\gamma^{2}(1-\delta)}\right]
    \frac{n_i(E_{\gamma}^{0},s,b,\ell)}{q} f_{\rm IC}(q,\delta)  
\label{eq:P_IC_G}    
\end{equation}
where the sum runs over all the considered low energy target photons. The Lorentz factor $\gamma = E_e/m_e$ and $\delta = E_{\gamma}/E_e$ and $E_{\gamma}^0$ is the energy for the initial target photon,
\begin{equation}
    E_{\gamma}^0 = \frac{m_{e}^2 E_{\gamma}}{4qE_{e}(E_{e}-E_\gamma )}, \quad f_{\rm IC}(q,x) = 2q\ln q+q+1-2q^2+\frac{x^{2}(1-q)}{2(1-x)}.
\end{equation}
The $e^\pm$ energy loss is dominated by IC scattering with background photons and synchrotron radiation due to intergalactic magnetic field, i.e., $b^{\rm G} =  b^{\rm G}_{\rm IC} + b^{\rm G}_{\rm syn}$. The IC energy loss is given by
\begin{equation}
b^{\rm G}_{\rm IC}(E_e,s,b,\ell) = \sum_i 3\sigma_{T} \int^{\infty}_{0}  \dd{\epsilon}  \epsilon  \int^1_{\frac{1}{4\gamma^2}}\dd{q} n_i(\epsilon,s,b,\ell)
\frac{(4\gamma^{2}-\Gamma_{\epsilon})q-1}{(1+\Gamma_{\epsilon}q)^3} f_{\rm IC}(q,-\Gamma_{\epsilon}q),
\label{eq:b_IC_G}
\end{equation}
with $\Gamma_{\epsilon}=4 \epsilon \gamma /m_e$. The electron energy loss due to synchrotron radition is 
\begin{equation}
b_{\rm syn}^{\rm G}(E_e,s,b,\ell) = \frac{4\sigma_T E_e^2}{3 m_e^2} \frac{B(s,b,\ell)^2}{2}
\label{eq:b_syn_G}
\end{equation}
We assume that the magnetic field is entirely composed of  a regular component thereby neglecting the presence of any  turbulent component\footnote{Adding a turbulent component of $2\,\mu$G changes our final bound by about $\mathcal{O}(1)\%$ for the DM masses where our bounds dominate. We have explicitly checked this for the $b\bar{b}$ channel.}. This implies that 
\begin{equation}
    B(r,z) = B_0\,{\rm exp}\left( - \frac{|r-r_\odot |}{r_B} - \frac{|z|}{z_B}\right)
\end{equation}
with $B_0 = 4.78\,\mu$G,   and $z = s \sin{b}$, $r_B = 10\,$kpc, $z_B = 2\,$kpc.

The $e^\pm$ diffusion halo function can be computed using diffusion equation in our Galaxy. We follow the formalism of Ref.\,\cite{Delahaye:2007fr}, where the solution of diffusion equation is written in terms of the Bessel and Fourier transform of the following form, assuming the $e^\pm$ diffusion (with diffusion length $\lambda_D$) happens within cylindrical diffusive halo with radius $R$ and vertical extension $z=-L$ to $z=+L$, beyond which $e^\pm$ escaped into the intergalactic medium. 
\begin{equation}
    \tilde{I}_\eta(\lambda_D,r,z) = \sum_{n,m=1}^{\infty} \vspace{-0.1cm}J_0\left(\frac{\alpha_n r}{R}\right) \sin{\left(\frac{m\pi (z+L)}{2L}\right)} \, e^{-\frac{\lambda_D^2}{4} \left(\left(\frac{m\pi}{2L}\right)^2 + \left(\frac{\alpha_n}{R}\right)^2 \right)}\,R^{n,m}_\eta,
    \label{eq:Itilde}
\end{equation}
where $J_i$ is the $i^{\rm th}$ order Bessel function of first kind. The zeros of $J_0$ are denoted by $\alpha_n$. The diffusion length is $\lambda_D = \sqrt{4 \kappa_0 \tau_E (E_e^{\delta-1}-(E^s_e)^{\delta-1})/(1-\delta)}$ with $\tau_E = 10^{16}\,$s and diffusion coefficient $\kappa = \kappa_0 (E/\GeV)^\delta$ with  normalisation $\kappa_0$. The Bessel and Fourier transform of relevant normalized DM density is given by\,\cite{Delahaye:2007fr}
\begin{equation}
    R^{n,m}_\eta = \frac{2}{J_1(\alpha_n)^2 R^2}\int_0^R \dd r^\prime J_0\left(\frac{\alpha_n r^\prime}{R}\right) \frac{1}{L}\int_{-L}^{+L} \dd z^\prime \sin{\left(\frac{m\pi (z^\prime+L)}{2L}\right)} \left(\frac{\rho_\chi(r^\prime)}{\rho_\odot}\right)^\eta
\label{eq:R_eta_nm}    
\end{equation}

\begin{figure*}[t]
\begin{center}
	\subfloat[\label{sf:I_IC_decay_l_Ee}]{\includegraphics[angle=0.0,width=0.5\textwidth]{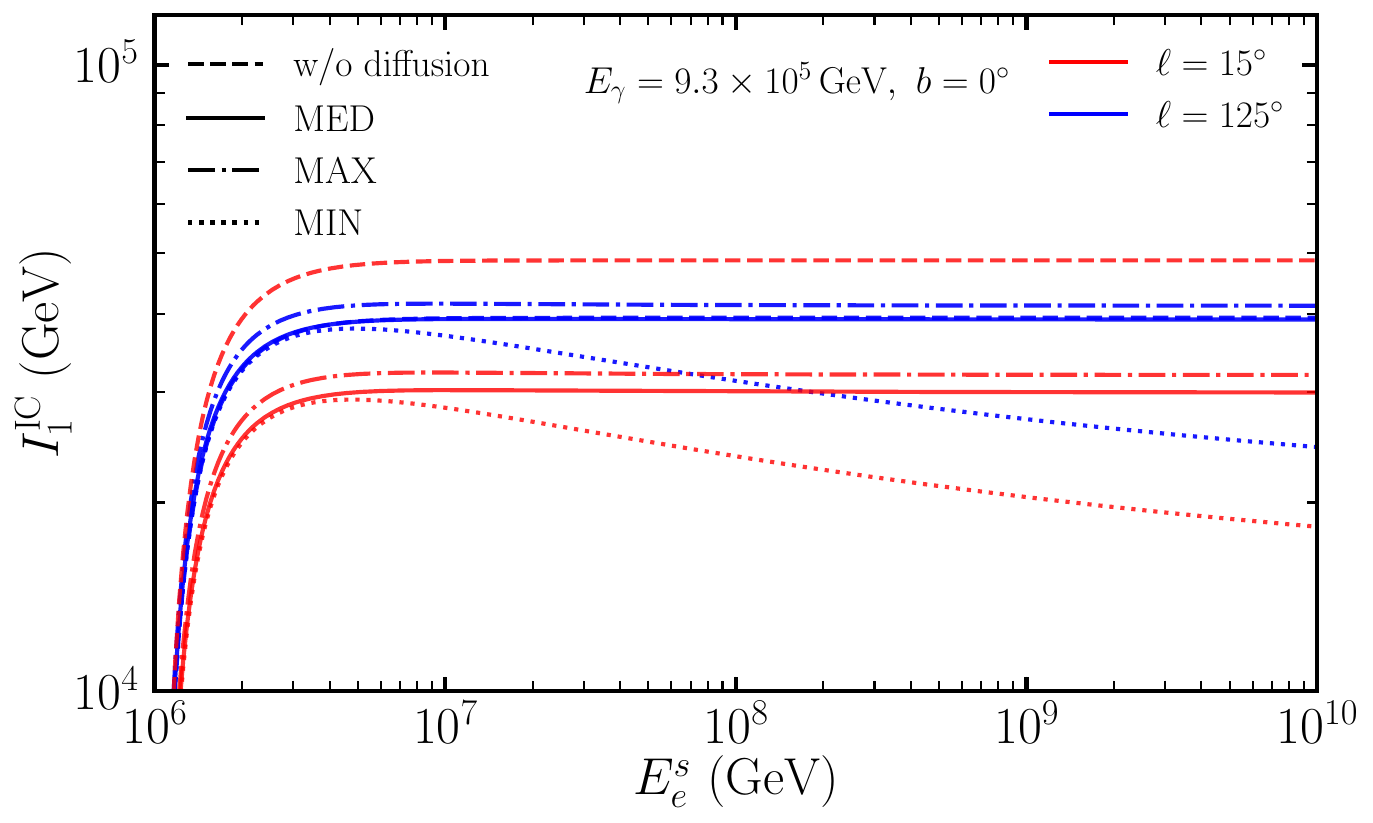}}~\subfloat[\label{sf:I_IC_decay_b_Ee}]{\includegraphics[angle=0.0,width=0.5\textwidth]{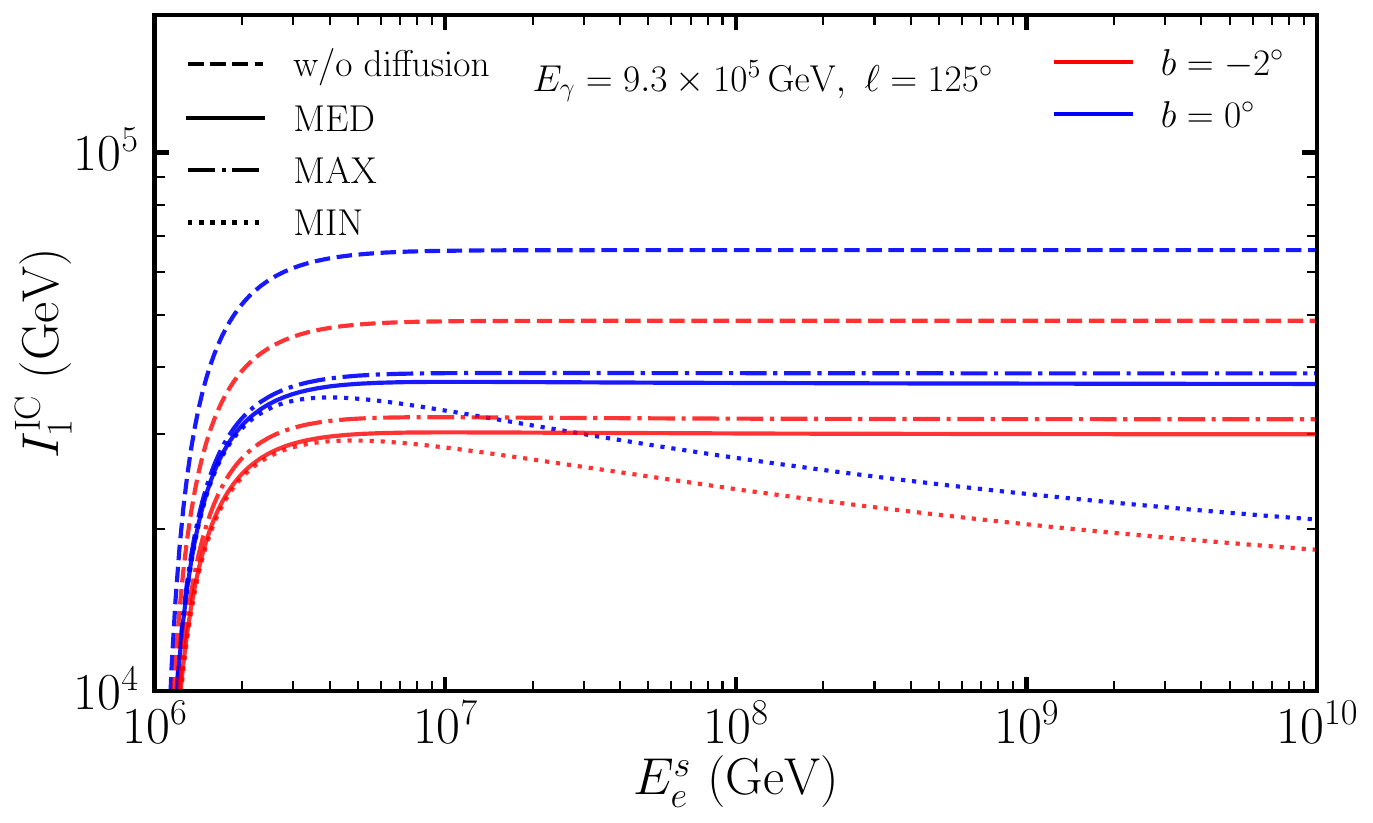}}~~\\	
	\caption{The IC halo function for decaying DM ($I_1^{\rm IC}$, see Eq.\,\eqref{eq:I_IC}) as a function of the injected electron energy $E_e^s$, for a fixed observed photon energy $E_{\gamma} = 9.3 \times 10^5\,\GeV$. The solid, dot-dashed, and dotted lines correspond to the MED, MAX, and MIN electron propagation models, respectively. Dashed lines represent the case where electron diffusion is neglected, i.e., $\tilde{I}_1 = \rho_\chi/\rho_\odot$. Panel~(a) shows the result for Galactic latitude $b = 0^\circ$, with red and blue lines corresponding to Galactic longitudes $\ell = 15^\circ$ and $\ell = 125^\circ$, respectively. Panel~(b) shows the variation for Galactic latitudes $b = -2^\circ$ and $b = 0^\circ$ by red and blue lines, with $\ell = 125^\circ$ fixed.}
	\label{fig:I_IC_Ee_decay}
\end{center}	
\end{figure*}

%$\delta=0.70$, $\kappa_0=0.0112\,$kpc$^2/$Myr and $L=4\,$kpc

\begin{figure*}[t]
\begin{center}
	\subfloat[\label{sf:I_IC_decay_l_Eg}]{\includegraphics[angle=0.0,width=0.5\textwidth]{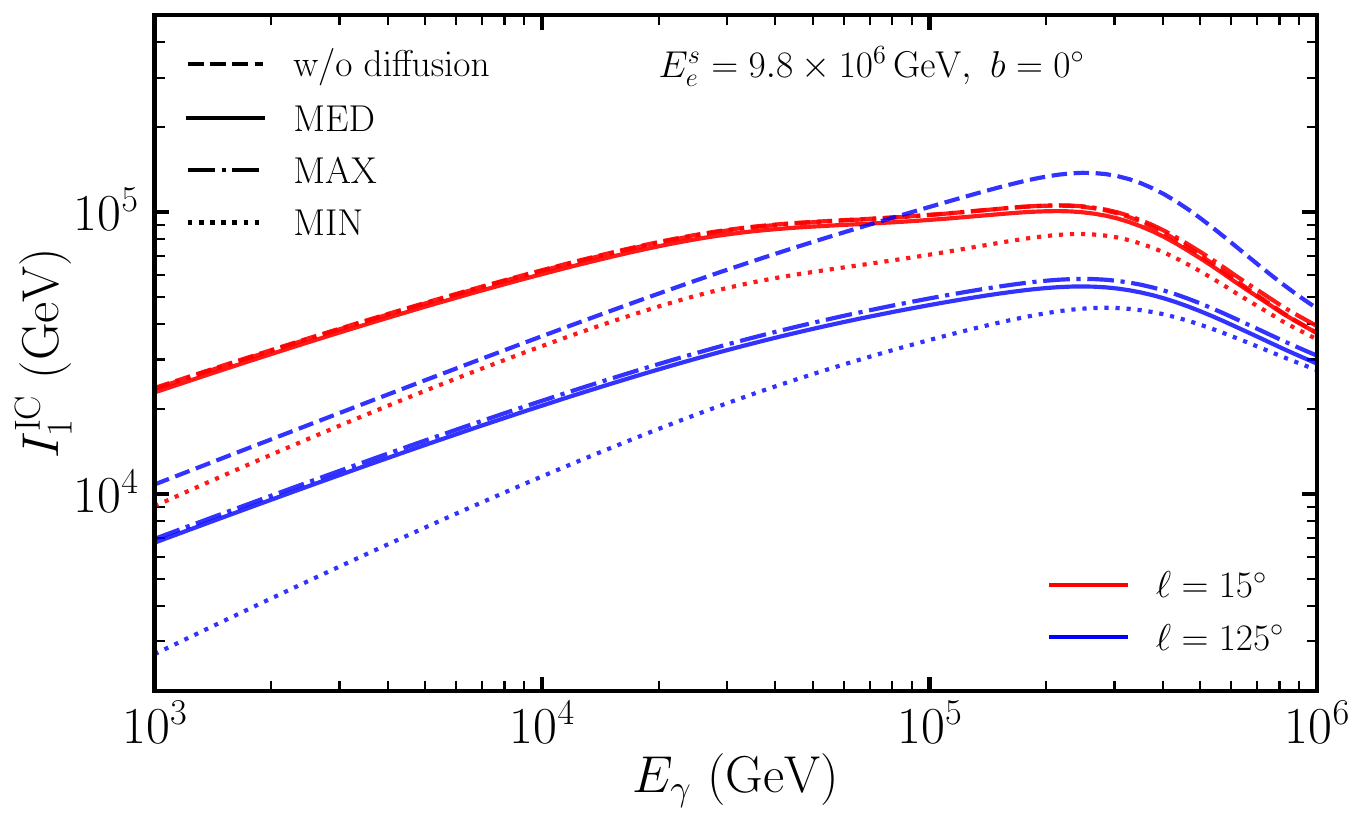}}~~
	\subfloat[\label{sf:I_IC_decay_b_Eg}]{\includegraphics[angle=0.0,width=0.5\textwidth]{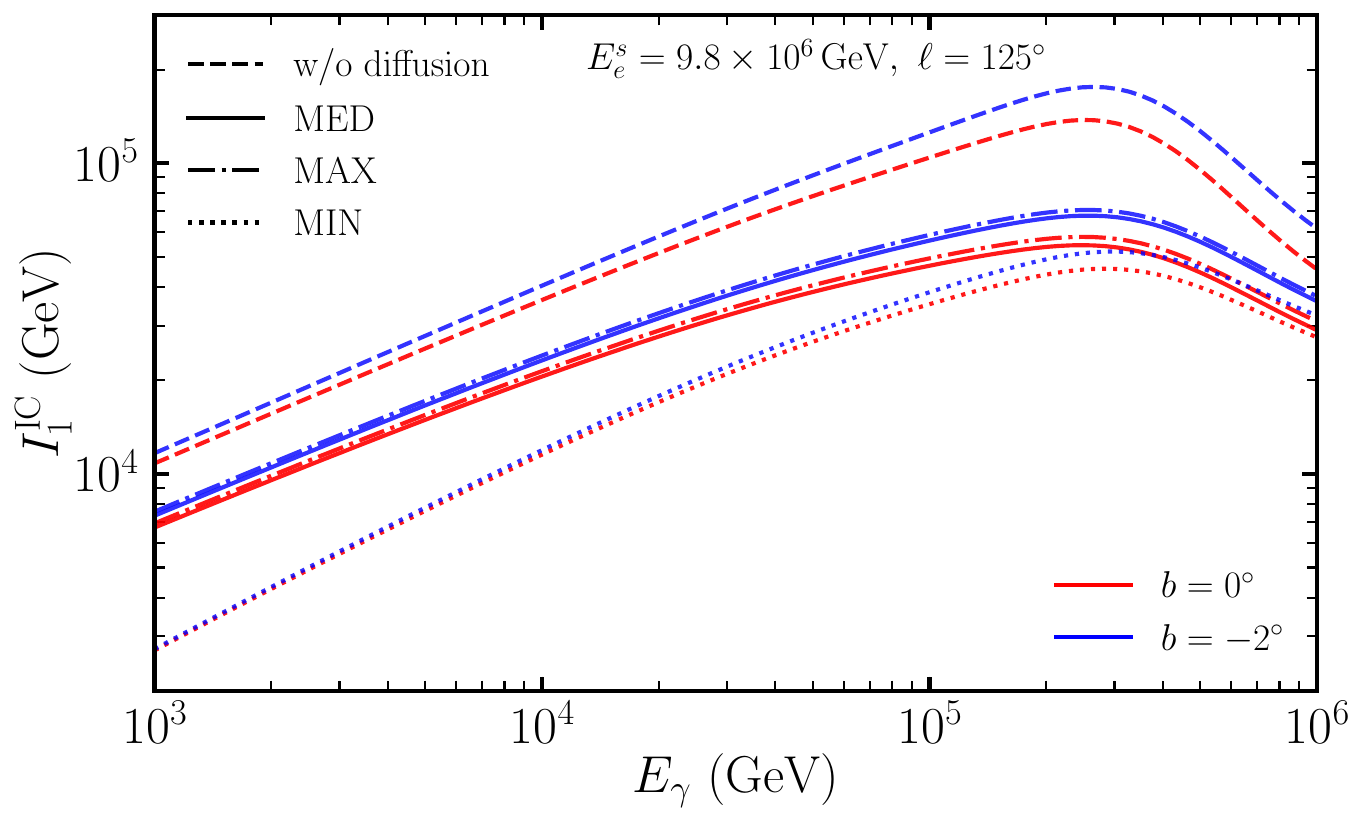}}~~\\	
	\caption{Same as Fig.\,\ref{fig:I_IC_Ee_decay} but for against photon energy $E_{\gamma}$ while keeping injected electron energy $E_e^s=9.8 \times 10^6\,\GeV$.}
	\label{fig:I_IC_Eg_decay}
\end{center}	
\end{figure*}

In our numerical computations, we adopt the MIN, MED, and MAX cosmic-ray propagation models, with parameters taken from Ref.\,\cite{Delahaye:2007fr}. These models have been updated in Ref.\,\cite{Genolini:2021doh} using AMS-02 data, however that does not affects our final results, as IC contribution is sub-dominant, as shown in appendix \ref{app:IC-flux}.  For DM  searches using high-energy $\gamma$-rays, spatial diffusion effects are often neglected \cite{Esmaili:2015xpa, Chianese:2019kyl, Leung:2023gwp, LHAASO:2022yxw}. To illustrate the impact of spatial diffusion, we display the IC halo function for decaying DM, $I_1^{\rm IC}$ for different  propagation models in Figs.\,\ref{fig:I_IC_Ee_decay} and \ref{fig:I_IC_Eg_decay}.

Fig.\,\ref{fig:I_IC_Ee_decay} presents $I_1^{\rm IC}$ as a function of the source electron energy $E_e^s$, for a fixed observed photon energy of $9.3 \times 10^5$\,GeV, across various Galactic coordinates $(b, \ell)$. At this photon energy, the dominant contribution to $P_{\rm IC}^{\rm G}$ arises from IC scattering off CMB photons. For fixed $E_\gamma$, the IC power $P_{\rm IC}^{\rm G}$ approximately scales as $E_e^{-2}$ (see Eq.\,\eqref{eq:P_IC_G}) above the relevant electron energy threshold, while the total energy loss ($b^{\rm G}$) rate scales as $E_e^2$. Given the relatively flat behavior of $\tilde{I}_1$ with respect to $E_e$ in the MED and MAX models \cite{Delahaye:2007fr} (where in our energy range electron diffusion length $\lambda_D$ remains smaller than the halo half-thickness $L$), the integrand $P_{\rm IC}^{\rm G}  \tilde{I}_1 / b^{\rm G}$ of $I_1^{\rm IC}$ (see Eq.\,\eqref{eq:I_IC})  scales as $E_e^{-4}$. This implies that the dominant contribution arises from lower-energy electrons, hence increasing the upper integration limit $E_e^s$ yields negligible changes. This behavior is reflected in the flat profile of $I^{\rm IC}_1$ for $E_e^s \gtrsim E_\gamma$ in the MED and MAX models in Fig.\,\ref{fig:I_IC_Ee_decay}. For the relevant electron energies in the MED and MAX models, the diffusion length does not exceed $4\,\mathrm{kpc}$ (the slab thickness used in the MED model), beyond which $\tilde{I}_1$ begins to differ between the two models. In contrast, for the MIN model, $\tilde{I}_1$ starts to decrease near $\lambda_D \sim 1\,\mathrm{kpc}$, as electrons with such diffusion lengths escape the Galactic diffusion zone into the intergalactic medium (IGM). This results in a visible suppression of $I^{\rm IC}_1$, as seen in the dotted curves in Fig.\,\ref{fig:I_IC_Ee_decay}. Variations in $I^{\rm IC}_1$ with Galactic coordinates $(b, \ell)$ are primarily driven by the spatial dependence of the DM density. For comparison, we also present results for a typical case where spatial diffusion is neglected. In this limit, $\tilde{I}_1$ in Eq.\,\eqref{eq:I_IC} is effectively replaced by $\rho_\chi/\rho_\odot$ for decaying DM. Comparing the solid, dotted, and dot-dashed curves of the same color with the diffusion-less result reveals that, in certain regions of the sky, propagation effects can alter $I^{\rm IC}_1$ by a factor of approximately 3.

In Fig.\,\ref{fig:I_IC_Eg_decay}, we plot $I_1^{\rm IC}$ as a function of $E_\gamma$, fixing the injected electron energy at $9.8 \times 10^6$\,GeV for different choices of Galactic coordinates $(b, \ell)$. At lower photon energies, the behavior of $I_1^{\rm IC}$ is primarily governed by the emitted power $P_{\rm IC}^{\rm G}$, which scales approximately linearly with $E_\gamma$ (see Eq.\,\eqref{eq:P_IC_G}). As $E_\gamma$ approaches the injected electron energy, $I_1^{\rm IC}$ is expected to decrease. However, for the chosen electron energy of $9.8 \times 10^6$\,GeV, this decline begins at lower $E_\gamma$ due to photon attenuation effects. Notably, for the selected values of $b$, smaller values of $\ell$ correspond to regions with higher densities of SL and IR backgrounds, leading to increased attenuation. The variation in the overall magnitude of $I_1^{\rm IC}$ across different $(b, \ell)$ values is primarily driven by changes in the DM density along the line of sight. As evident from the plots, propagation effects can modify the IC electron halo function by a factor of $\sim 2$. We observe similar behavior for annihilating DM, with the key difference that the electron halo function $\tilde{I}_2$ scales as $(\rho_\chi/\rho_\odot)^2$.

%For a fixed choice of $(b, \ell)$, attenuation causes the shape of $I_1^{\rm IC}$ to become nearly identical for $E_e^s \gtrsim 10^7$\,GeV.
%This results in a flattening of $I_1^{\rm IC}$ above $E_\gamma \gtrsim 3 \times 10^4$\,GeV (see red lines in Fig.\,\ref{sf:I_IC_decay_l_Eg}), up to the point where attenuation becomes dominant.
% propagated electron spectrum is given by
% \begin{equation}
%     \psi_\eta(E_e,s,b,\ell) = \frac{\mathcal{K}}{b(E_e)}\int_{E_e} \dd E_e^s \, \tilde{I}(E_e^s,E_e,s,b,\ell) \dv{N_e}{E_e^s},
% \end{equation}

%%%%%%%%%%
\subsection{Extragalactic photons from DM}
\label{subsec:galactic}
%%%%%%%%%
Having discussed the galactic photon flux from DM, we now turn to the EG contribution. Similar to the galactic case, EG photons can be produced promptly from DM decay or indirectly via IC scattering, where $e^\pm$ from DM decay scatter off low-energy photons from the CMB. Additionally, as previously mentioned, photons with energies above a few TeV originating from EG sources undergo attenuation due to interactions with background photons. The resulting $e^\pm$ pairs can subsequently scatter off CMB or EBL photons  and produce high-energy photons, a process known as electromagnetic cascade. We discuss these contributions below.

The EG prompt contribution is given by 
\begin{equation}
\frac{\mathrm{d}^2 \phi_\gamma}{\mathrm{d} E_\gamma \, \mathrm{d} \Omega}\bigg\rvert_{\rm EG}^{\rm p}  = \frac{1}{4 \pi \eta} \int ^{\infty}_{0}\frac{\dd{z}}{H(z)\left(1+z\right)^3} \left(\frac{\rho^{\rm E}_{\chi}(z)}{ m_{\chi}}\right)^\eta \mathcal{P^\prime}_\eta^{\gamma} e^{-\tau^{\rm EG}_{\gamma \gamma} (E_{\gamma},z)} \mathcal{B}_\eta^{\rm EG},
\label{eq:EG-prompt}
\end{equation}
with the Hubble parameter $H(z)$, and the cosmological DM density evolving as $\rho^{\rm E}_{\chi}(z) = \rho_0 (1+z)^3$, and $\rho_0$ is the present day DM density. The DM decay or annihilation-dependent factor, ${\mathcal{P}^\prime}_\eta^{\gamma}$, follows from Eq.\,\eqref{eq:p_etaDM}; however, the photon spectrum $\dd{N_\gamma}/\dd{E_\gamma}$ must now be evaluated at $E_\gamma(1+z)$, corresponding to a photon observed today with energy $E_\gamma$. The attenuation of high-energy EG photons due to pair production is incorporated through the optical depth $\tau^{\rm EG}_{\gamma \gamma}$. The EG boost factor \cite{Ullio:2002pj}, $\mathcal{B}_\eta^{\rm EG}$, is equal to 1 for decaying DM, while its expression for annihilating DM will be discussed later.

The EG photon opacity is given by
\begin{equation}
\tau^{\rm EG}_{\gamma \gamma}(E_{\rm \gamma},z) = \sum_i \frac{3 \sigma_T m_e^4}{4 E_{\gamma}^2} \int_0^z \dd{z^\prime}  \frac{1}{H(z^\prime) (1+z^\prime)^3} \int_0^{\infty} \dd{\epsilon} \frac{\mathcal{S} (s_{z^\prime})
}{\epsilon^2} n_i(\epsilon,z), 
\label{eq:EG-att}
\end{equation}
The differential cosmological number density of background photons is denoted by $n_i(\epsilon,z)$, which depends on both redshift $z$ and photon energy $\epsilon$. We include contributions from both the CMB and from EBL, adopting the model from Ref.\,\cite{Saldana-Lopez:2020qzx}. Note that EBL models typically provide the comoving intensity, $I_\epsilon$, which can be converted to the differential number density via $n(\epsilon, z) = 4\pi (1+z)^3 I_\epsilon/c\epsilon^2$. The function $\mathcal{S}(s_{z^\prime})$, with $s_{z^\prime} = E_\gamma \epsilon (1+z^\prime)/m_e^2$, arises from the angular integration of the pair production cross section and is given by
\begin{eqnarray}
\mathcal{S}(s_{z^\prime}) &=& \frac{1}{6 s_{z^\prime}}\Big[  - \pi^2 s_{z^\prime}  + 6 \sqrt{(s_{z^\prime} - 1)s_{z^\prime}}  - 12 \sqrt{(s_{z^\prime} - 1) s_{z^\prime}^3} - 3 s_{z^\prime}\ln(s_{z^\prime}) + 6 s_{z^\prime} \ln^2 2   \nonumber \\ 
&+& 3 s_{z^\prime} \ln^2f^{-}(s_{z^\prime}) -9 s_{z^\prime} \ln f^{+}(s_{z^\prime}) + 6 s_{z^\prime}^2 \ln f^{+}(s_{z^\prime})  - 3 s_{z^\prime} \ln^2 f^{+}(s_{z^\prime})  \nonumber \\ 
&+& \ln f^{-}(s_{z^\prime}) \left( 
    -3 + 3 s_{z^\prime} - 6 s_{z^\prime}^2 
    - 12 s_{z^\prime} \ln2
    + 6 s_{z^\prime} \ln f^{+}(s_{z^\prime})
\right)  \nonumber \\ 
&+& 3 \ln f^{+}(s_{z^\prime}) + 12 s_{z^\prime} \operatorname{Li}_2 \left(1 -  \left(f^{-}(s_{z^\prime})/2\right)   \right) \Big]\Theta(s_{z^\prime} - 1).
\label{eq:S}
\end{eqnarray}
where $f^{\pm}(s_{z^\prime}) = 1\pm \sqrt{(s_{z^\prime}-1)/s_{z^\prime}}$ and $\Theta$ is the Heaviside theta function and the Spence function is
$$\operatorname{Li}_2 (u) =  \int^u_1 \dd{t} \frac{\ln t}{1-t}.$$
This numerically matches with relevant part of Ref.\,\cite{Biteau:2015xpa}.

Assuming a NFW halo, we compute the EG boost factor following\,\cite{Ando:2013ff,Ng:2013xha}
\begin{equation}
    \mathcal{B}_2^{\rm EG} = \left(\frac{1}{\rho^{\rm E}_\chi(z)}\right)^2 \int \dd M_{\rm host} \dv{N}{M_{\rm host}}\left(1+b_{\rm sh}(M_{\rm host})\right) \int \dd V \rho_{\rm host}^2\,.
\end{equation}
Here, $M_{\rm host}$ denotes the host halo mass, and the halo mass function, $\dd N/\dd M_{\rm host}$, is adopted from the \texttt{hmf} package\,\cite{Murray:2013qza}, assuming a conservative minimum halo mass of $10^6\,M_\odot$. It is worth noting that the minimum halo mass depends on the kinetic decoupling temperature of DM; for WIMP-like DM, it can be as small as $\sim 10^{-12}\,M_\odot$\,\cite{Bringmann:2009vf}. The subhalo boost factor, $b_{\rm sh}$, is calculated using the fitting function given in Ref.\,\cite{Ando:2019xlm} with the concentration-mass relation, $c_{\rm v}$ from Ref.\,\cite{Correa:2015dva}. Assuming NFW profile for the host halo, we obtain
\begin{equation}
   \int \dd V \rho_{\rm host}^2 = \frac{4 \pi \tilde{r}_s^3 \tilde{\rho}_s^2}{3}\left(1-\frac{1}{1+c_{\rm v}}\right), \quad \tilde{\rho}_s = \frac{M_{\rm host}}{4 \pi \tilde{r}_s^3 }\left(\ln(1+c_{\rm v})-\frac{c_{\rm v}}{1+c_{\rm v}}\right)^{-1}.
\end{equation}
The scale density and scale radius of the halo are denoted by $\tilde{\rho}_s$ and $\tilde{r}_s$, respectively. The scale radius is related to the concentration parameter via $\tilde{r}_s = r_{\rm v}/c_{\rm v}$, where $r_{\rm v}$ is the virial radius. The host halo mass is given by $M_{\rm host} = 4 \pi r_{\rm v}^3 \Delta \rho_{\chi}(z)/3$, with $\Delta = 200$. The EG boost factor depends on redshift $z$ and can be as large as $\sim 10^5$ for late universe. Inclusion of the subhalo boost factor ($b_{\rm sh}$) can further enhance it by an additional factor of $\sim 2$.

The EG IC contribution is given by 
\begin{eqnarray}
\frac{\mathrm{d}^2 \phi_\gamma}{\mathrm{d} E_\gamma \, \mathrm{d} \Omega}\bigg\rvert_{\rm EG}^{\rm IC}  = \frac{1}{2\pi E_{\gamma} \eta }\int\limits^{\infty}_{0}\frac{\dd{z} e^{-\tau^{\rm EG}_{\gamma \gamma}}\mathcal{B}_\eta^{\rm EG}}{H(z)(1+z)^3}   \hspace{-0.2cm}  \int\limits^{m_{\chi} \eta /2}_{E_{\gamma}(1+z)} \hspace{-0.3cm} \dd{E_e} \frac{P^{\rm EG}_{\rm IC}(E_{\gamma},E_e,z)}{b^{\rm EG}_{\rm IC}(E_e,z)}\hspace{-0.2cm}\int\limits^{m_{\chi} \eta /2}_{E_{e}} \hspace{-0.3cm} \dd{E^s_e}\mathcal{P}_\eta^e (E^s_e) 
\label{eq:IC_EG}
\end{eqnarray}
where, $\mathcal{P}_\eta^e(E_e^s)$ follows the same form as in Eq.\,\eqref{eq:p_etaDM}, but evaluated for $e^\pm$ energy $E_e^s$. The EG IC power transferred from an electron to a photon, $P^{\rm EG}_{\rm IC}$, and the EG energy loss rate, $b^{\rm EG}_{\rm IC}$, both arise from IC scattering with CMB photons. Including the EBL in the computation of $P^{\rm EG}_{\rm IC}$ and $b^{\rm EG}_{\rm IC}$ has a numerically negligible effect. However, in the case of attenuation, characterized by the optical depth $\tau^{\rm EG}_{\gamma \gamma}$ given in Eq.\,\eqref{eq:EG-att}, the EBL plays a significant role due to its higher photon energies, which allow for pair production at lower values of $E_\gamma$ compared to the CMB. In Eq.\,\eqref{eq:IC_EG}, we neglected the propagation of $e^\pm$, as they are expected to scatter with CMB photons before diffusing over cosmological distances. The EG IC power is obtained by replacing $E_\gamma$ in Eq.\,\eqref{eq:P_IC_G} with $E_\gamma (1+z)$ and noting that the background CMB photon density changes with redshift. A similar modification applies to the energy loss due to IC scattering. The energy loss contribution from synchrotron radiation is negligible, given the expected strength of the EG magnetic field.

In addition to the EG prompt and EG IC contributions, we also include the cascaded component using the {\tt $\gamma-$Cascade} library\,\cite{Capanema:2024nwe}, adopting the EBL model from Ref.\,\cite{Saldana-Lopez:2020qzx}. For a given two-body final state, the injected electron spectrum is computed using \texttt{HDMSpectrum}, and then convolved iteratively with the IC photon spectrum provided by {\tt $\gamma-$Cascade} to obtain the cascaded spectrum. 

For heavy decaying DM ($m_\chi \gtrsim 10^6\,$GeV), the spectral shape of the EG $\gamma$-ray flux becomes nearly universal below $E_\gamma \lesssim 10^6\,$GeV. This can be understood by noting that as the DM mass increases, the number density decreases as $1/m_\chi$, while the total power injected into photons increases linearly with $m_\chi$\,\cite{Cohen:2016uyg}. These effects nearly cancel each other, leading to an approximately universal spectrum, except near the high-energy endpoint. However, this behavior does not hold for annihilating DM. In this case, the annihilation rate scales as $1/m_\chi^2$, while the injected power scales as $2 m_\chi$, resulting in a flux that decreases with mass. As a consequence, the spectrum is not universal but acquires a DM mass-dependent normalization.

%%%%%%%%%%%%%%
\section{Results}
\label{sec:results}
\begin{figure*}[t]
\begin{center}
	\subfloat[\label{sf:flux_decay_b}]{\includegraphics[angle=0.0,width=0.5\textwidth]{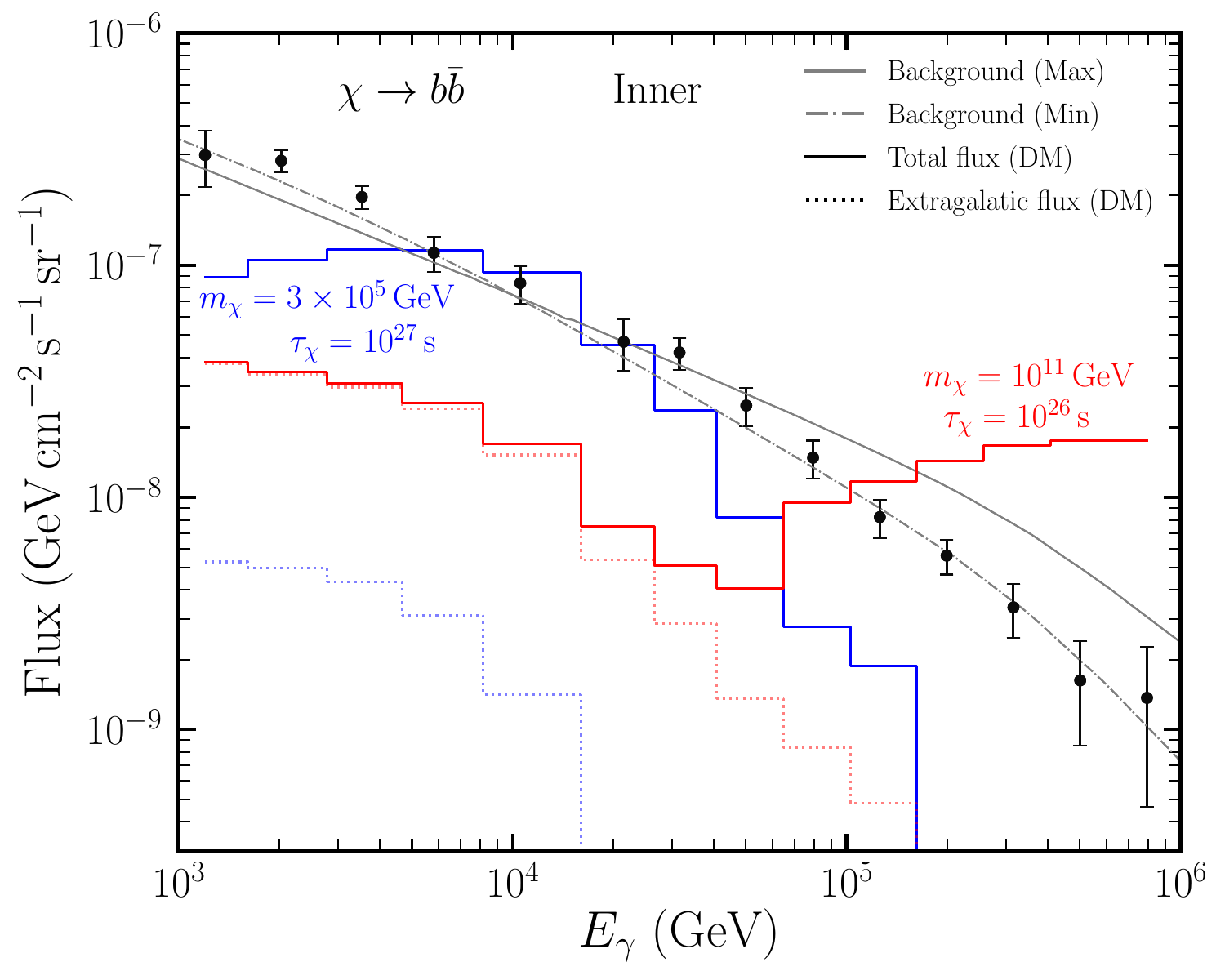}}~
	\subfloat[\label{sf:flux_annihilation_b}]{\includegraphics[angle=0.0,width=0.5\textwidth]{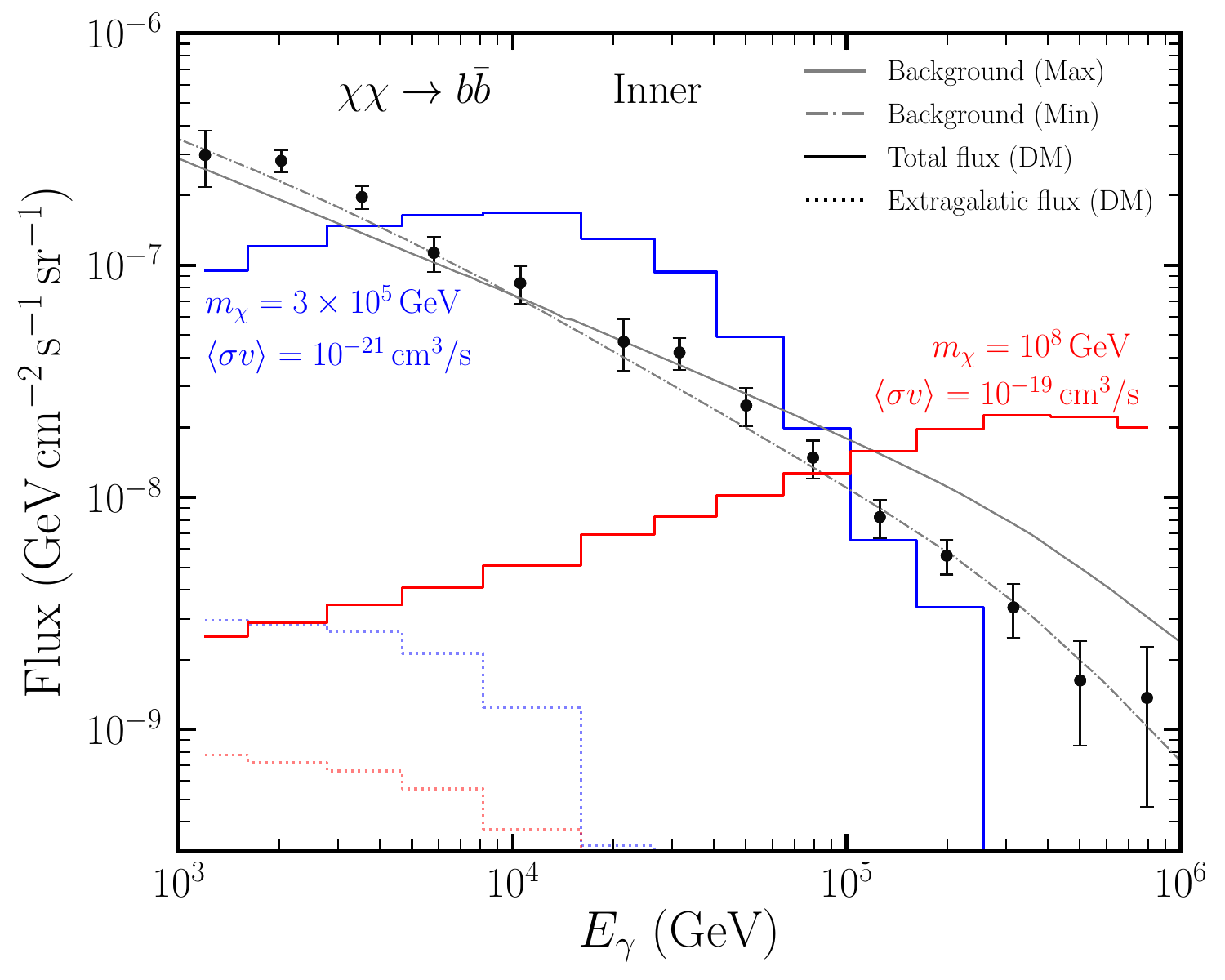}}
	\caption{The photon flux from DM for the inner region ($15^\circ < \ell < 125^\circ,\, |b|<5^\circ $) of the Galactic plane is shown. The red and blue solid lines represent the total flux, obtained by summing the Galactic and EG contributions for the labeled DM parameters. For the Galactic IC flux, we assume the MED propagation model, while comparisons with other propagation models are presented in Fig.\,\ref{fig:flux_ic_compare}. The EG contribution is shown by the dotted lines. The corresponding observed LHAASO flux, along with uncertainties, is indicated by black points. Two choices of background models labeled as Background\,(Min) and background\,(Max) from Ref.\,\cite{DeLaTorreLuque:2025zsv} are displayed by the gray dot-dashed and solid lines respectively. The left panel (a) corresponds to DM decaying to $b\bar{b}$, whereas the right panel (b) shows DM annihilating to $b\bar{b}$, including both Galactic and EG boost factors.}
	\label{fig:flux}
\end{center}	
\end{figure*}

%%%%%%%%%%%%%
Having discussed the formalism to compute the photon flux from DM annihilation and decay, we now present the corresponding results and the methodology adopted to search for DM using LHAASO data. In Fig.\,\ref{fig:flux}, we show the photon flux from DM decay and annihilation into $b\bar{b}$ in the left and right panels, respectively, for the inner ($15^\circ < \ell < 125^\circ,\, |b|<5^\circ $) galactic plane region of the sky. The solid blue and red lines represent the total DM induced flux, obtained by summing the galactic and EG contributions for the labelled DM masses and parameters. For the galactic component, the IC flux carries some uncertainty arising from the propagation of $e^\pm$. However, this contribution remains subdominant in the context of DM searches using LHAASO data, as IC contribution itself is subdominant which we discussed in the appendix~\ref{app:IC-flux}.  The total EG flux is shown by the corresponding dotted lines in Fig.\,\ref{fig:flux}. For a decaying DM mass of $3 \times 10^5\,$GeV, the EG flux is mostly subdominant in the displayed photon energy range. In contrast, for a DM mass of $10^{11}\,$GeV, the EG contribution dominates at lower photon energies, as high-energy photons from heavy DM decay are efficiently reprocessed via electromagnetic cascades, as illustrated by the red dotted line in Fig.\,\ref{sf:flux_decay_b}\footnote{We note that our EG IC flux for decaying DM, computed using the full Klein–Nishina cross section and redshift dependence as given in Eq.\,\eqref{eq:IC_EG}, differs by a factor of $\sim 1.5$ from the PPPC results \cite{Cirelli:2010xx, Buch:2015iya}, for the same EBL model as in PPPC.}. This results in an enhancement of the total flux at lower energies, as indicated by the red solid line in Fig.\,\ref{sf:flux_decay_b}. The flux from annihilating DM is shown in Fig.\,\ref{sf:flux_annihilation_b}, where we have included the boost factor for both the galactic and EG components. As evident from the comparison of solid and dotted lines in Fig.\,\ref{sf:flux_annihilation_b}, the EG contribution to the total flux itself remains subdominant even after including large EG boost factor.

With the photon flux computed from DM, we now proceed to discuss the data used to search for DM signals. In Fig.\,\ref{fig:flux}, we show the observed diffuse $\gamma$-ray flux from LHAASO for the inner Galactic plane region, as measured using both the KM2A and WCDA arrays. The black solid points represent the LHAASO data along with the uncertainty where statistical and systematic uncertainties added in quadrature. This data is adopted from Ref.\,\cite{LHAASO:2024lnz}. The dominant component of the observed flux arises from CR interactions in the ISM. There are, however, some observational uncertainties in the CR proton flux at energies above $\sim 10^6\,$GeV, particularly due to discrepancies between the measurements by IceTop and KASCADE. To account for this uncertainty in the predicted $\gamma$-ray background, we use two CR background models in our analysis, referred to as ``Min'' and ``Max'' set up (these should not be confused with electron propagation models MIN and MAX) in Ref.\,\cite{DeLaTorreLuque:2025zsv}. These models are shown in Fig.\,\ref{fig:flux} as gray solid and dot-dashed lines, and are labelled as Background\,(Min) and Background\,(Max), respectively. It is worth noting that both background models are consistent with the LHAASO measurements and can simultaneously fit the diffuse $\gamma$-ray observations from Fermi-LAT at lower energies\,\cite{DeLaTorreLuque:2025zsv}.

Using the LHAASO data, background models, and representative DM signals shown in Fig.\,\ref{fig:flux}, we employ the following $\chi^2$ statistic to search for DM:
\begin{equation}
    \chi^2 = \frac{\left({\rm data } - {\rm background} - {\rm DM\,signal}\right)^2}{\sigma_{\rm data}^2}
    \label{eq:chisquare}
\end{equation}
where, the data are represented by the black points in Fig.\,\ref{fig:flux}, while the background can be chosen as either of the gray lines shown in the same figure. The uncertainty in the data is denoted by $\sigma_{\rm data}$. The predicted signal from DM depends on the DM lifetime or annihilation cross section, its mass, and the final state into which the DM decays or annihilates. For a given DM mass and decay/\,annihilation channel,   we vary $\tau_\chi$/\,$\langle \sigma v \rangle$ until the change in $\chi^2$, $\Delta \chi^2 = \chi^2 - \chi^2_{\rm min}$, reaches 2.71, which corresponds to a $95\%$ confidence level limit on the $\tau_\chi$/\,$\langle \sigma v \rangle$. Here, $\chi^2_{\rm min}$ refers to the minimum $\chi^2$ value obtained for the best-fit parameters, given a particular background model, DM interaction type, and mass.

\begin{figure*}[t]
\begin{center}
	\subfloat[\label{sf:limit_decay_b}]{\includegraphics[angle=0.0,width=0.5\textwidth]{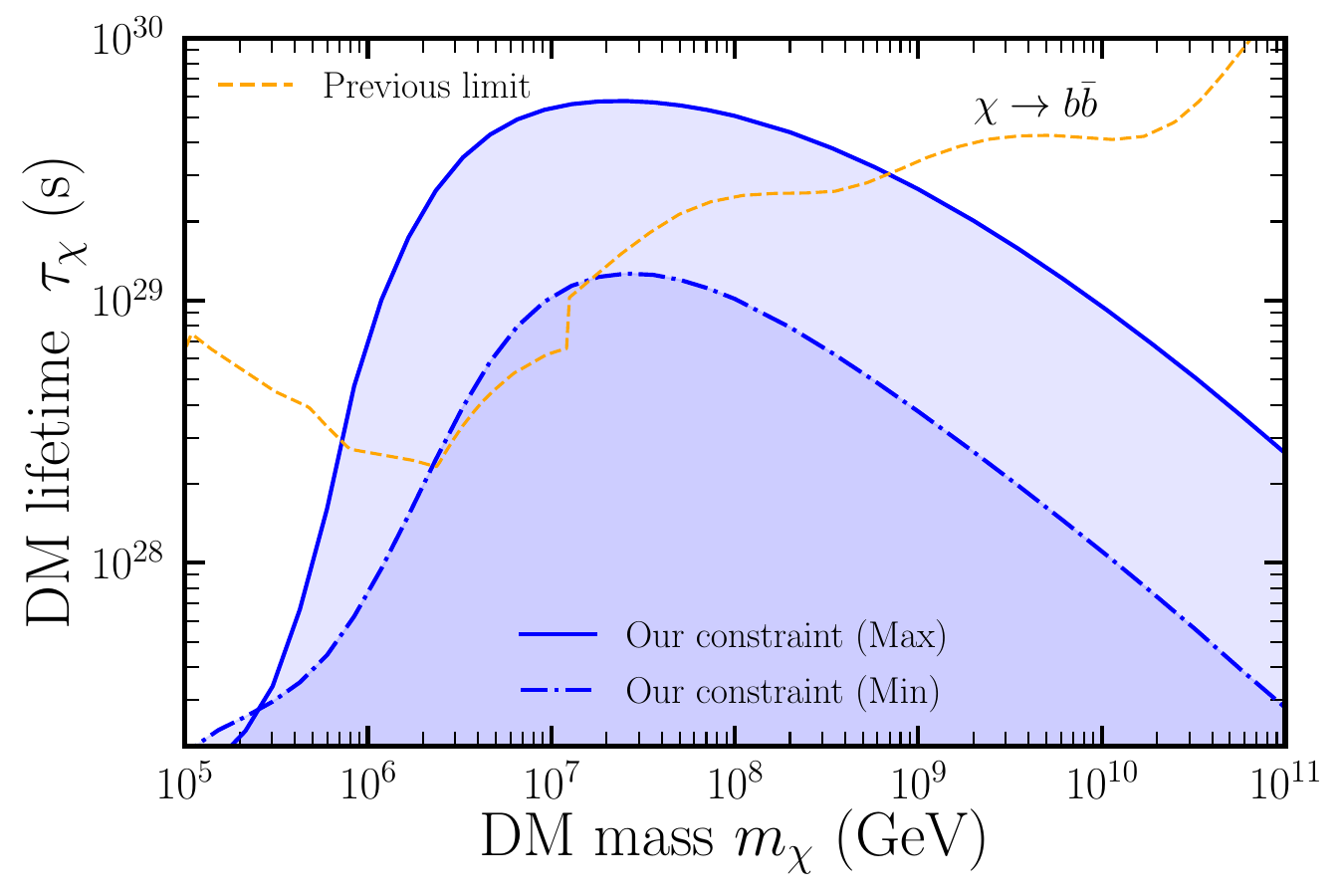}}~
	\subfloat[\label{sf:limit_annihilation_b}]{\includegraphics[angle=0.0,width=0.5\textwidth]{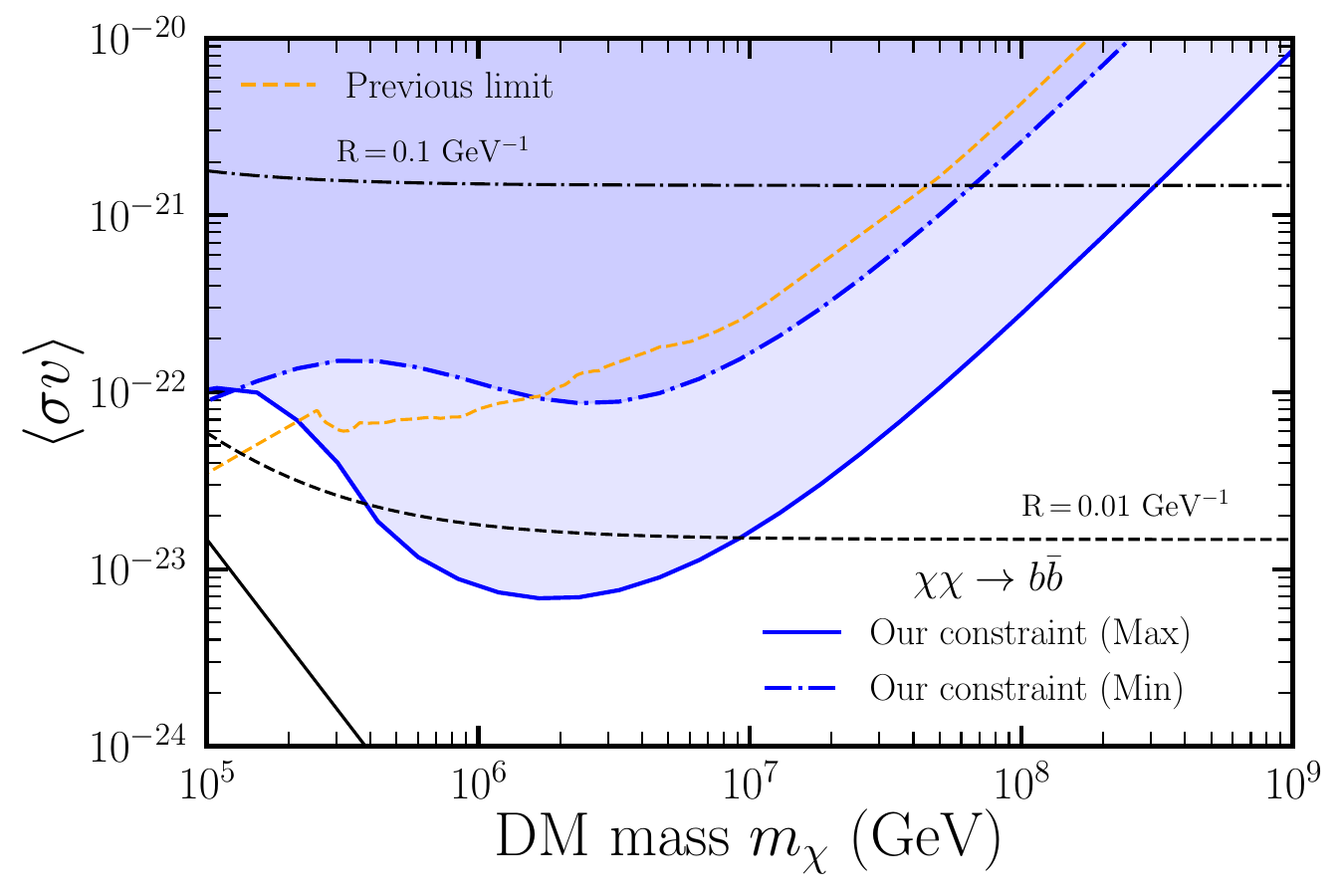}}
	\caption{Bounds on the DM parameter space. The orange solid line shows existing limits. Our bounds, assuming the Min and Max background models (see gray lines in Fig.\,\ref{fig:flux}, not to be confused with electron propagation models), are shown by the blue dot-dashed and solid lines, respectively, ruling out the blue shaded regions of the DM parameter space. In the left panel (a), we consider DM decaying to $b\bar{b}$. In the right panel (b), DM is assumed to annihilate to $b\bar{b}$, where we also show the unitarity bound for point-like DM by the black solid line, and for composite DM with labeled radius $R$, by the black dashed and dot-dashed lines.}
	\label{fig:limit}
\end{center}	
\end{figure*}

The results of the above analysis are shown in Fig.\,\ref{fig:limit} for DM decaying and annihilating into $b\bar{b}$ in the left and right panels, respectively. Although our analysis also includes data from the outer sky region ($125^\circ < \ell < 235^\circ,\, |b|<5^\circ $), we do not display the corresponding limits, as the constraints obtained from the outer sky region are generally weaker. This is primarily due to the higher DM density in the inner sky region compared to the outer region. In both panels of Fig.\,\ref{fig:limit}, we show the combined previous limits from Refs.\,\cite{Cohen:2016uyg, Blanco:2018esa, Bhattacharya:2019ucd, Ishiwata:2019aet, Chianese:2021jke, IceCube:2022clp, LHAASO:2024lnz,  Aloisio:2025nts} for decaying DM and from Refs.\,\cite{IceCube:2022clp, LHAASO:2024upb, Acharyya:2023ptu} for annihilating DM by the orange dashed lines. Our limits for the Min and Max (note to be confused with $e^\pm$ propagation model) background models are shown by the solid and dot-dashed blue lines, respectively, with the blue shaded regions indicating the excluded parameter space. 

The maximum photon energy that DM can produce increases with DM mass, while the overall flux decreases due to dilution in DM number density. However, this dilution does not fully offset the reduction in flux associated with higher photon energies, resulting in a strengthening of the limits as the DM mass increases at lower masses. This trend continues until the DM mass becomes large enough to contribute significantly to the highest energy bin in Fig.\,\ref{fig:flux}. Beyond this point, the DM-induced photon flux decreases, particularly at the highest photon energies, leading to a gradual weakening of the constraints. The Max background model typically yields stronger limits than the Min model because, in some energy bins, it predicts $\gamma$-ray fluxes that slightly exceed the observed data (while remaining consistent within $\sim 2\sigma$). This leaves less room for additional DM contributions and therefore results in tighter bounds. Conversely, in the lowest energy bins, the Min background model predicts higher $\gamma$-ray fluxes than the Max model, thereby providing slightly stronger constraints in lower DM mass regime.
\begin{figure*}[t]
\begin{center}
	\subfloat[\label{sf:limit_decay_tau}]{\includegraphics[angle=0.0,width=0.5\textwidth]{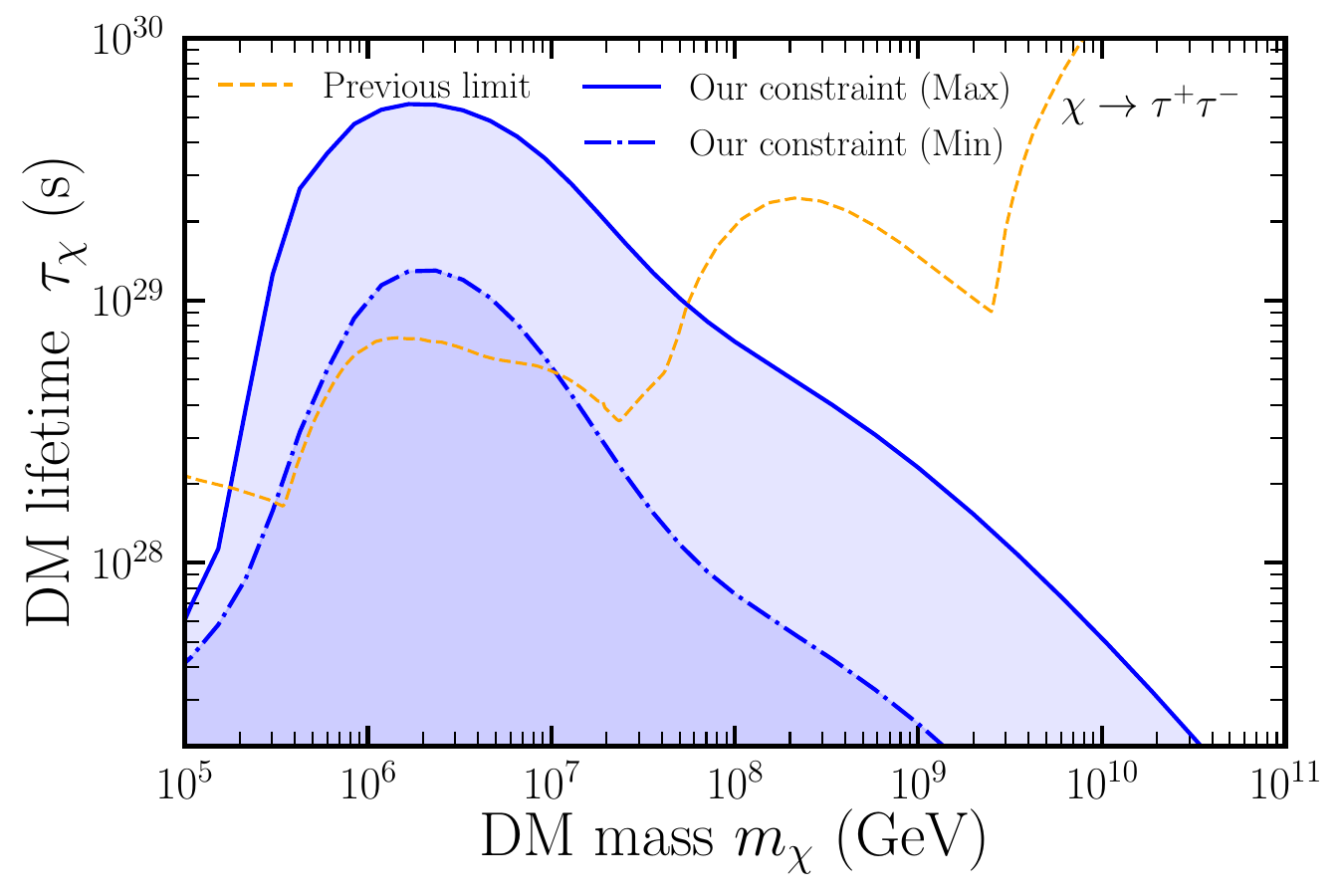}}~
	\subfloat[\label{sf:limit_annihilation_tau}]{\includegraphics[angle=0.0,width=0.5\textwidth]{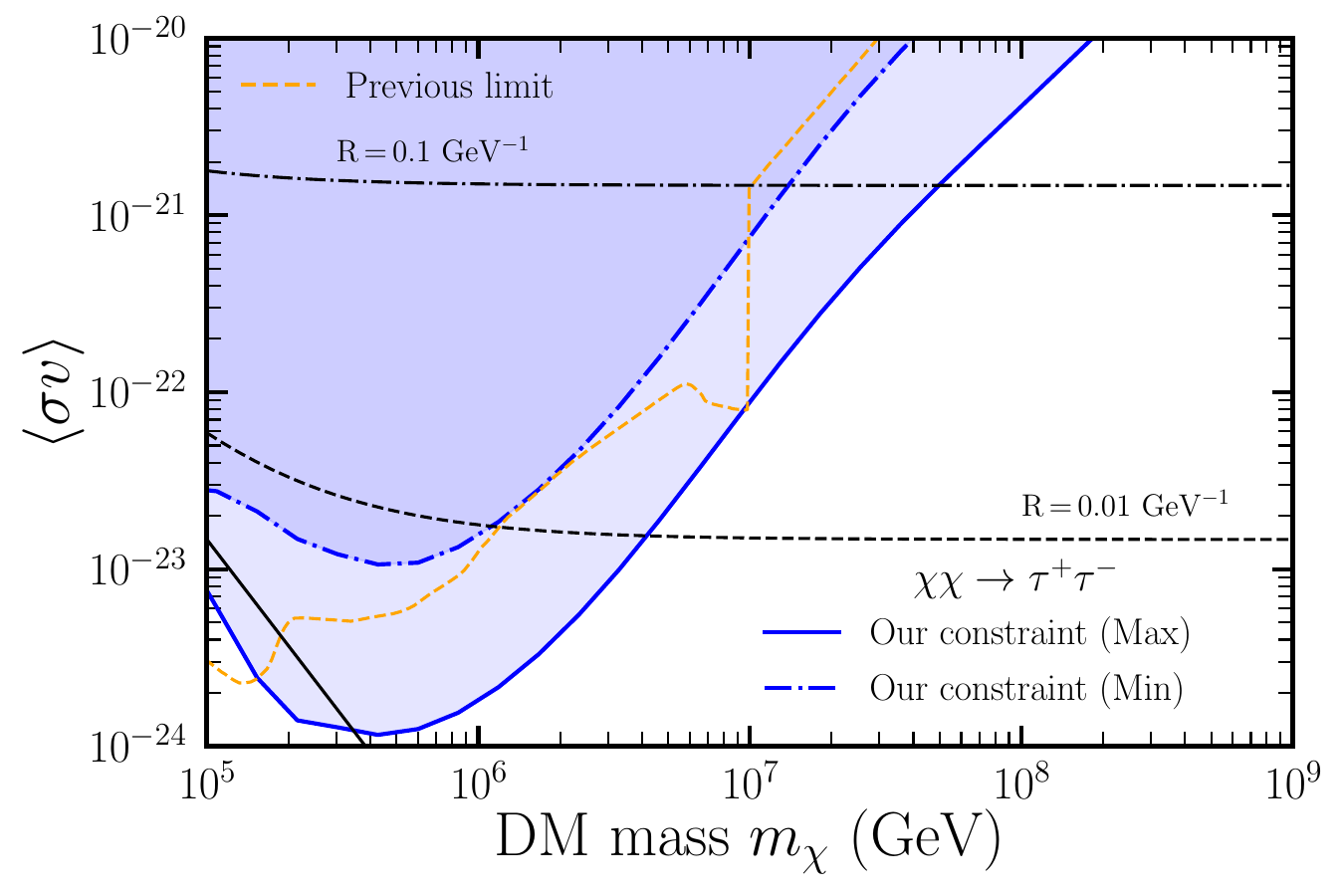}}
	\caption{Same as Fig.\,\ref{fig:limit} but for $\tau^+ \tau^-$ final states.}
	\label{fig:limit_tau}
\end{center}	
\end{figure*}

The simplest annihilating DM within the mass and annihilation cross section range shown in Fig.\,\ref{sf:limit_annihilation_b} violates the partial-wave unitarity bound \cite{Griest:1989wd}. There are several ways through which this limit can be softened \cite{Cirelli:2018iax, Frumkin:2022ror, Kramer:2020sbb, Kim:2019udq, Tak:2022vkb}. We illustrate one such case where the DM, instead of being a point particle, is a composite object with radius $R$. In this scenario, the partial-wave unitarity limit is maintained within the region $\langle \sigma v \rangle \leq 4 \pi (1+ m_\chi R v)^2/m_\chi^2 v$ \cite{Tak:2022vkb} with $v$ is the DM relative velocity, which we assume to be $10^{-3}$ in our analysis. For $R=0$, the corresponding upper limit is shown by the black solid line in Fig.\,\ref{sf:limit_annihilation_b}, whereas for $R=0.1\,\GeV^{-1}$ and $R=0.01\,\GeV^{-1}$ are depicted by the black dashed and dot-dashed lines, respectively. 

By comparing the blue lines with the orange dashed lines in Fig.\,\ref{fig:limit}, it is evident that our limits, regardless of the chosen background model, provide the strongest constraints in certain regions of the DM parameter space. The limits for other considered two-body final states, both for decaying and annihilating DM, are shown in Figs.\,\ref{fig:limit_decay_other} and \ref{fig:limit_annihilation_other}, respectively. As expected, photon-rich final states (e.g., $\tau^+ \tau^-$) yield stronger constraints than photon-poor states (e.g., $\nu_e \bar{\nu}_e$). Furthermore, among photon-rich final states, those producing harder photon spectra (e.g., $\tau^+ \tau^-$) provide tighter limits compared to softer photon final states (e.g., $e^+ e^-$). Generally, for the Min background model, we obtain stronger limits than previous results in some regions of the DM parameter space, except for the $\mu^+ \mu^-$, $\nu_e \bar{\nu}_e$, and $\nu_\mu \bar{\nu}_\mu$ final states. In contrast, for the Max background model, we find the strongest constraints across a larger region of parameter space for most final states. These results highlight the promising potential of LHAASO to probe and scrutinize heavy annihilating and decaying DM.
%%%%%%%%%
\section{Conclusion}
\label{sec:conclusion}
%%%%%%%%%
%
Identifying the nature of DM  remains one of the most pressing questions in modern science. The allowed DM mass range spans nearly $90$ orders of magnitude. In this paper, we focus on searching for heavy particle DM in the mass range $10^5$–$10^{11}$ GeV, which is challenging to probe using collider and direct detection experiments. For the first time, we utilized LHAASO observations of diffuse high-energy $\gamma$-rays from the Galactic plane, to search for annihilating and decaying DM signatures.

We focus on the two-body decay or annihilation of DM into SM final states, which generically produce high-energy photons from both Galactic and EG sources. We systematically compute both contributions. Within the Galactic component, photons arise promptly from DM decay or annihilation. Additionally, $e^\pm$ produced in these interactions can upscatter background photons—originating from the CMB, EBL, SL, and IR radiation—via IC scattering, generating secondary high-energy photons. For DM searches in the PeV energy regime, we incorporate, for the first time, the effect of electron propagation using the Bessel-function formalism. Since LHAASO’s sky coverage excludes regions with high SL and IR photon densities, the IC contribution is subdominant in the total flux, and $e^\pm$ propagation effects have a limited impact on our final constraints. We also include the EG contribution, accounting for prompt, IC, and cascade photons. In both Galactic and EG cases, attenuation due to photon pair production is a critical factor, and we incorporate it in our analysis. High-energy photons traveling cosmological distances are significantly attenuated, producing electron-positron pairs that subsequently interact with low-energy CMB and EBL photons via IC scattering, generating cascaded secondary photons. These cascade contributions are also included in our numerical calculations.

Having systematically included all these contributions, we next search for high-energy photons produced from DM in the latest LHAASO data. Our analysis accounts for DM-induced photons in the presence of the guaranteed background flux arising from CR interactions in the ISM. Although there are uncertainties in the prediction of this high-energy photon background, we adopt two well-motivated CR models that not only fit the LHAASO observations but also explain the diffuse gamma-ray flux measured by Fermi-LAT. Using these two background models, we do not find any conclusive evidence for DM. Instead, we set the strongest constraints on certain regions of the DM parameter space, independent of the considered background model, for most of the considered two-body final states. Given LHAASO’s expected multi-decade operational timeline and its transformative role in high-energy astrophysics, our results highlight its potential to play a pivotal role in uncovering the particle nature of DM.

\appendix
%%%%%%%%%%%%%%%%%%%%%
\section{IC flux comparison}
\label{app:IC-flux}
\begin{figure*}[h]
\begin{center}
	\subfloat[\label{sf:flux_decay_b_IC_compare}]{\includegraphics[angle=0.0,width=0.5\textwidth]{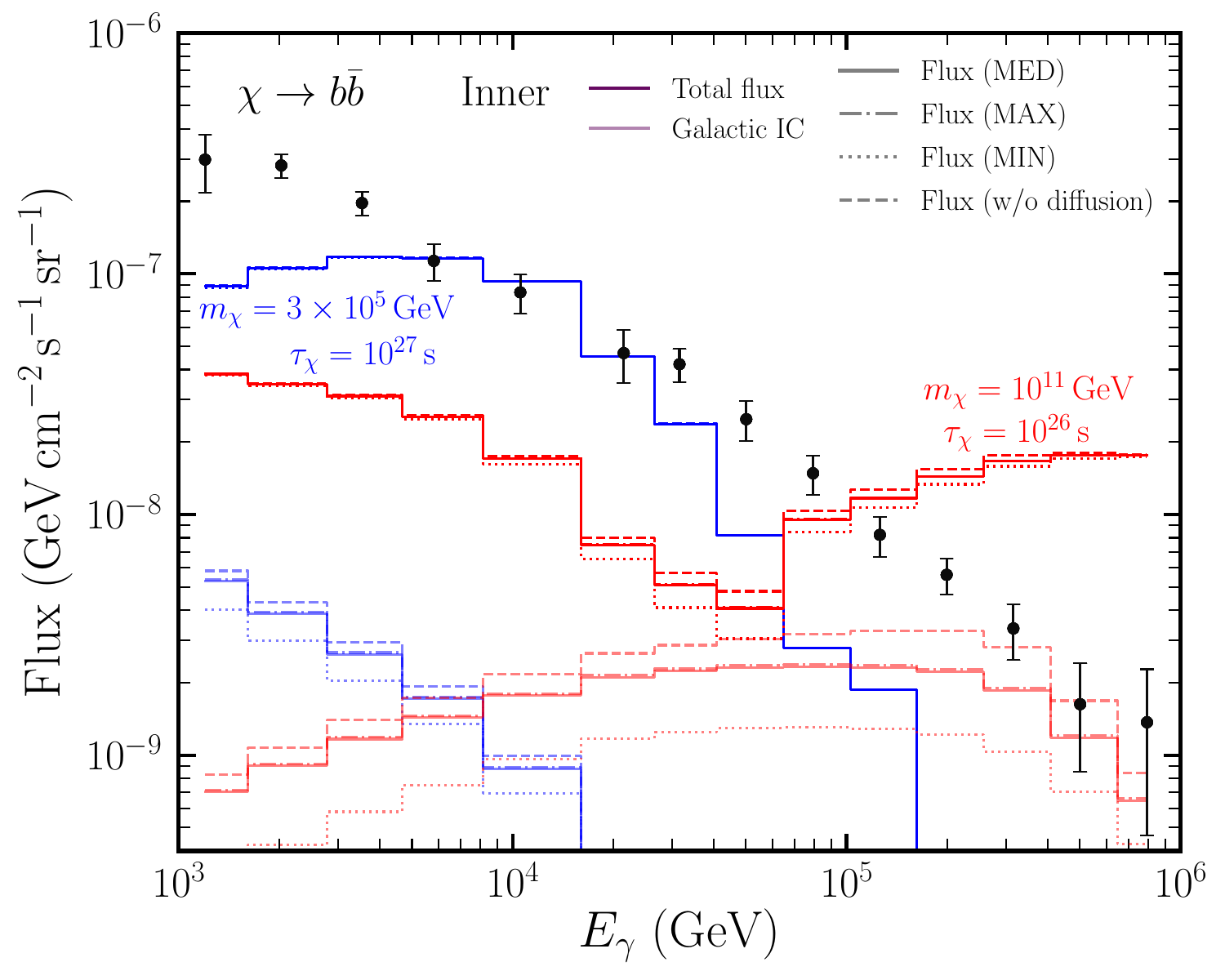}}
	\subfloat[\label{sf:flux_annihilation_b_IC_compare}]{\includegraphics[angle=0.0,width=0.5\textwidth]{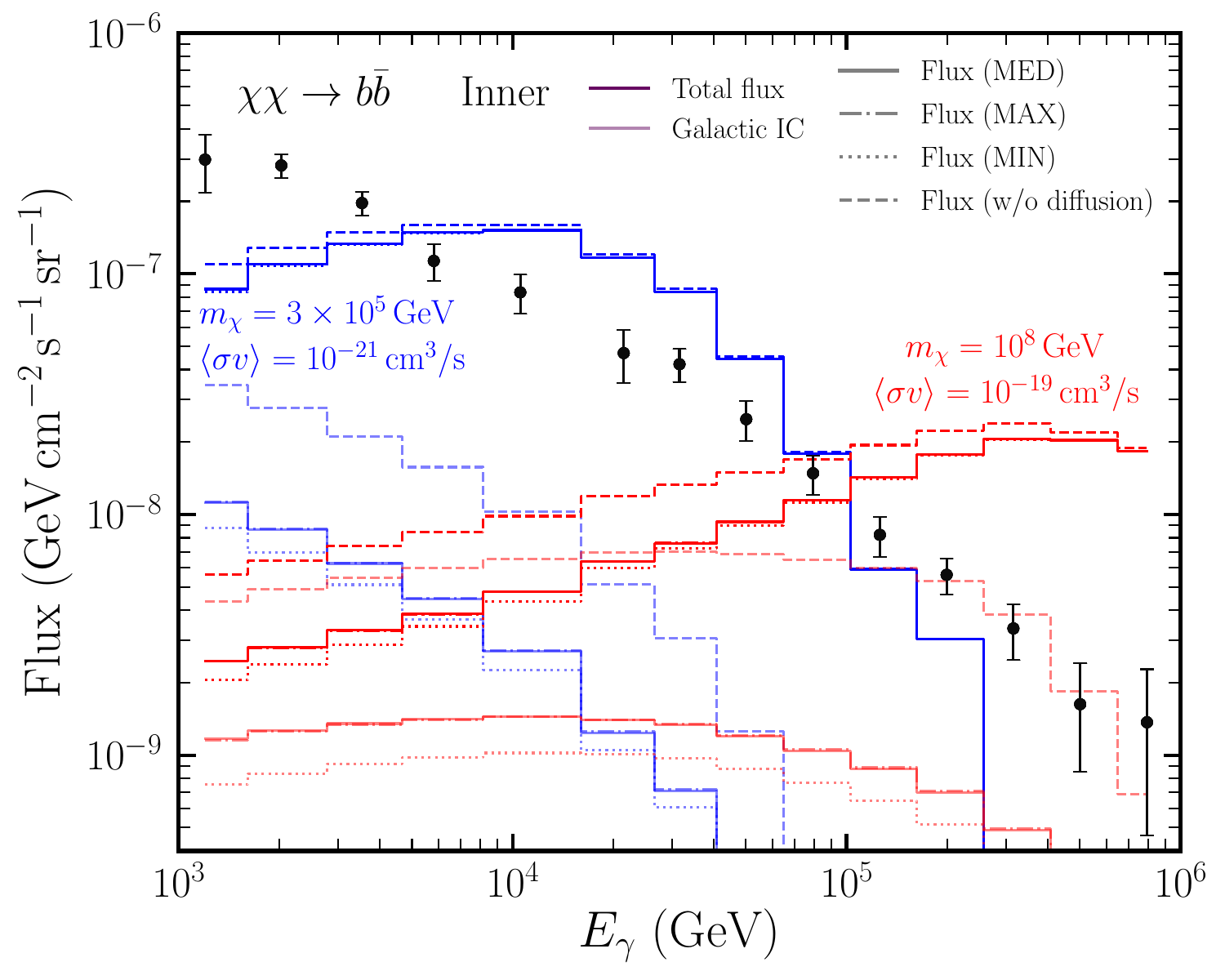}}
	\caption{DM induced photon flux and its comparison with the Galactic IC flux for different electron propagation models. The total flux for the MED, MAX, and MIN models is shown by the solid, dot-dashed, and dotted lines, respectively. The dashed line represents the flux for the case without diffusion, i.e. $\tilde{I}_\eta = (\rho_\chi/\rho_\odot)^\eta$ in Eq.\,\eqref{eq:IC_EG}. The corresponding lower-opacity lines indicate the contribution from the Galactic IC component.  As can be seen, regardless of the propagation model, the contribution from the Galactic IC flux remains subdominant and thus affects our bounds only minimally. Other details are the same as in Fig.\,\ref{fig:flux}.}
	\label{fig:flux_ic_compare}
\end{center}	
\end{figure*}
In this appendix, we discuss the impact of Galactic $e^\pm$ propagation on the DM-induced $\gamma$-ray flux. This is illustrated in Fig.\,\ref{fig:flux_ic_compare}. The total fluxes for the MIN, MED, and MAX propagation models are shown by the solid, dot-dashed, and dotted lines, respectively. The flux computed without considering electron diffusion, i.e., $\tilde{I}_\eta = (\rho_\chi/\rho_\odot)^\eta$ in Eq.\,\eqref{eq:IC_EG}, is shown by the dashed lines, this is the typical choice adopted in the literature\,\cite{Esmaili:2015xpa, Chianese:2019kyl, Leung:2023gwp, LHAASO:2022yxw}. The corresponding IC contributions are shown by the respective lower-opacity lines. While the IC flux itself may vary by a factor of a few, however the comparison with total flux shows that in the photon energy bins relevant for setting limits on DM parameters using the LHAASO data, the Galactic IC contribution remains numerically subdominant. As a result, it affects our bounds $\lesssim \mathcal{O}(10)\%$ level for heavier DM masses. Therefore, throughout the text, we present the results assuming the MED propagation model for $e^\pm$.    

\section{Decaying DM: other final states }
\label{app:decayingDM_other}
\begin{figure*}[h!]
\begin{center}
	\subfloat[\label{sf:limit_decay_u}]{\includegraphics[angle=0.0,width=0.31\textwidth]{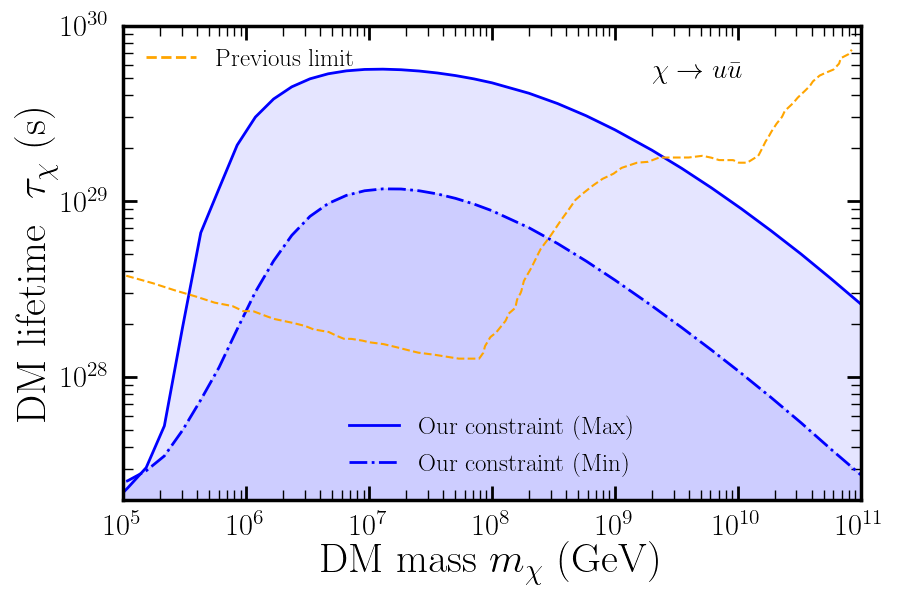}}~~
	\subfloat[\label{sf:limit_decay_d}]{\includegraphics[angle=0.0,width=0.31\textwidth]{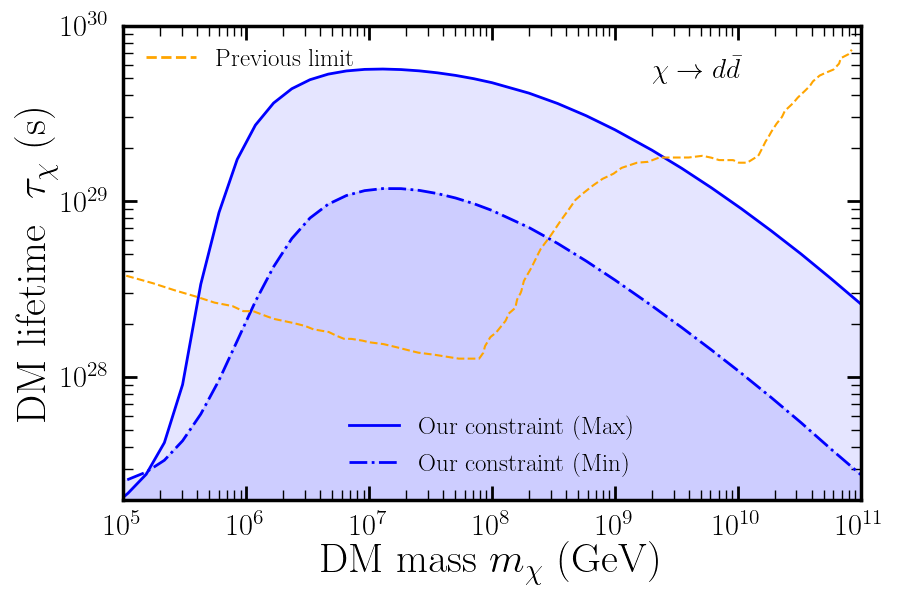}}~~
    \subfloat[\label{sf:limit_decay_c}]{\includegraphics[angle=0.0,width=0.31\textwidth]{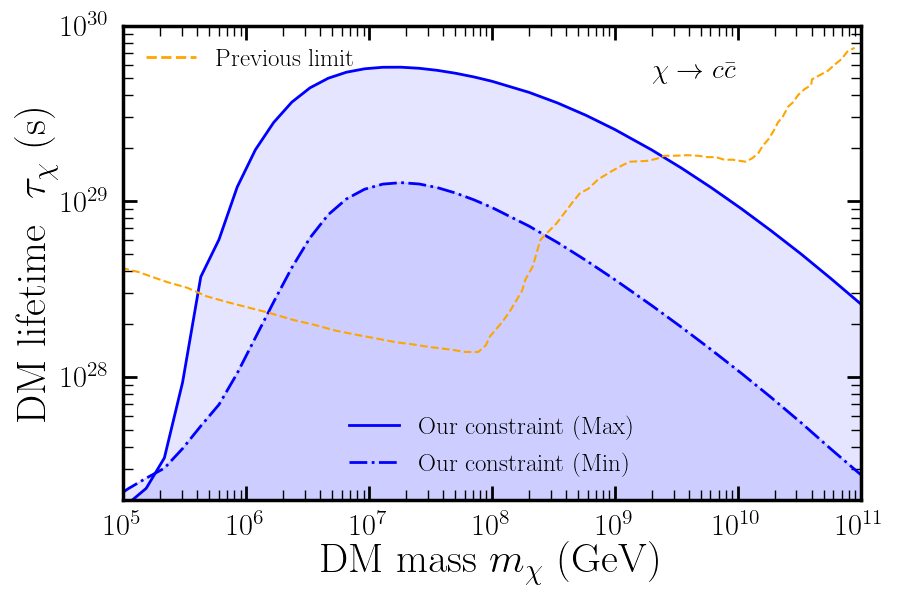}}~~ \\
	\subfloat[\label{sf:limit_decay_s}]{\includegraphics[angle=0.0,width=0.31\textwidth]{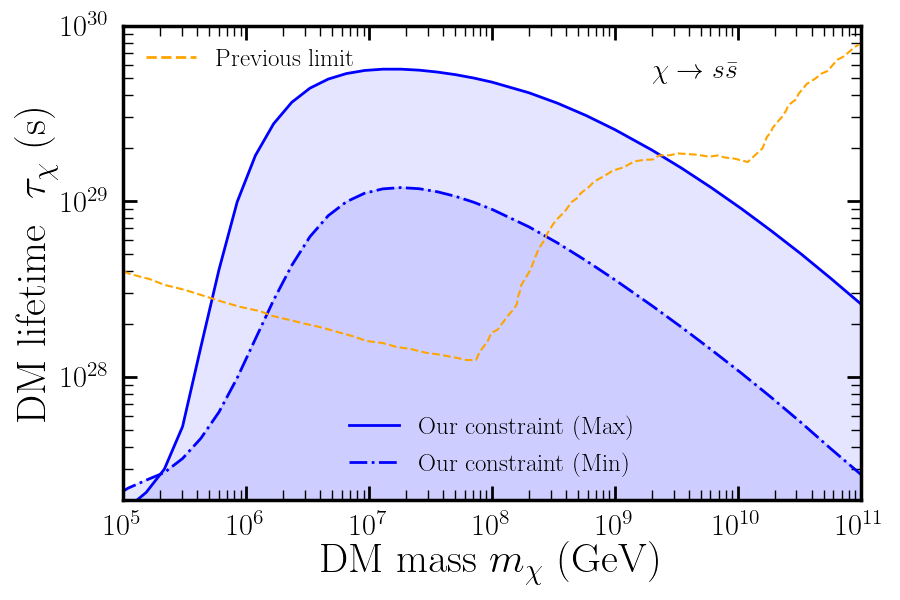}}~~
	\subfloat[\label{sf:limit_decay_t}]{\includegraphics[angle=0.0,width=0.31\textwidth]{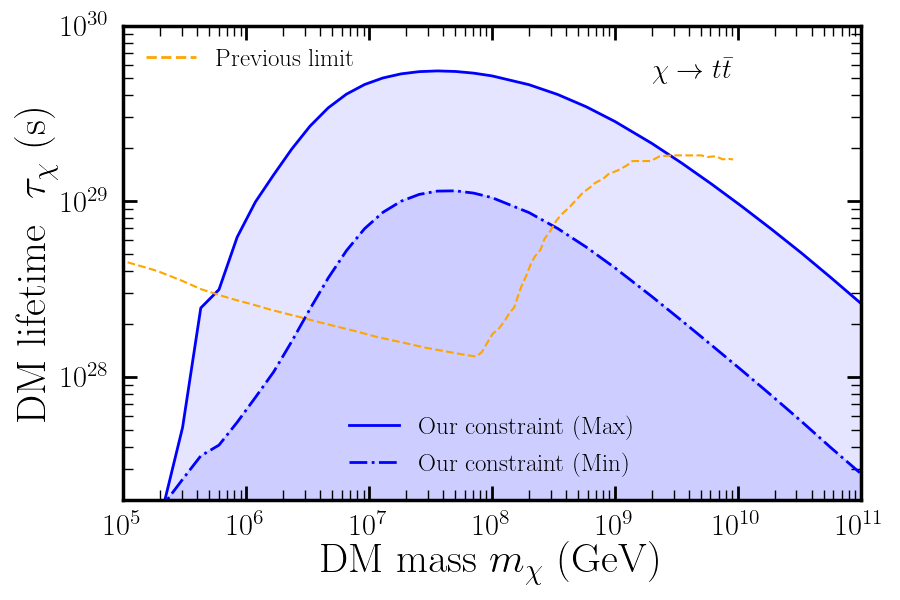}}~~
    \subfloat[\label{sf:limit_decay_e}]{\includegraphics[angle=0.0,width=0.31\textwidth]{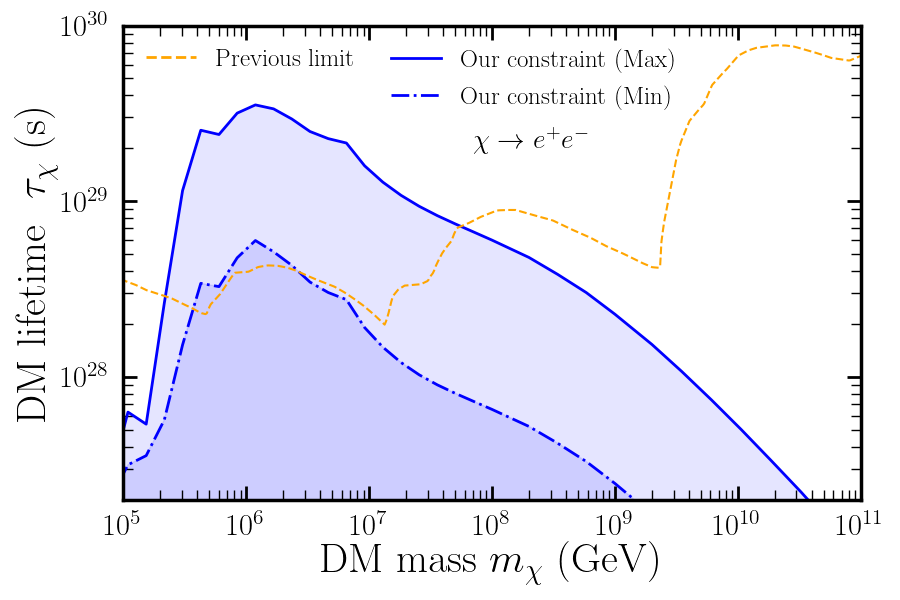}}~~ \\  
    \subfloat[\label{sf:limit_decay_mu}]{\includegraphics[angle=0.0,width=0.31\textwidth]{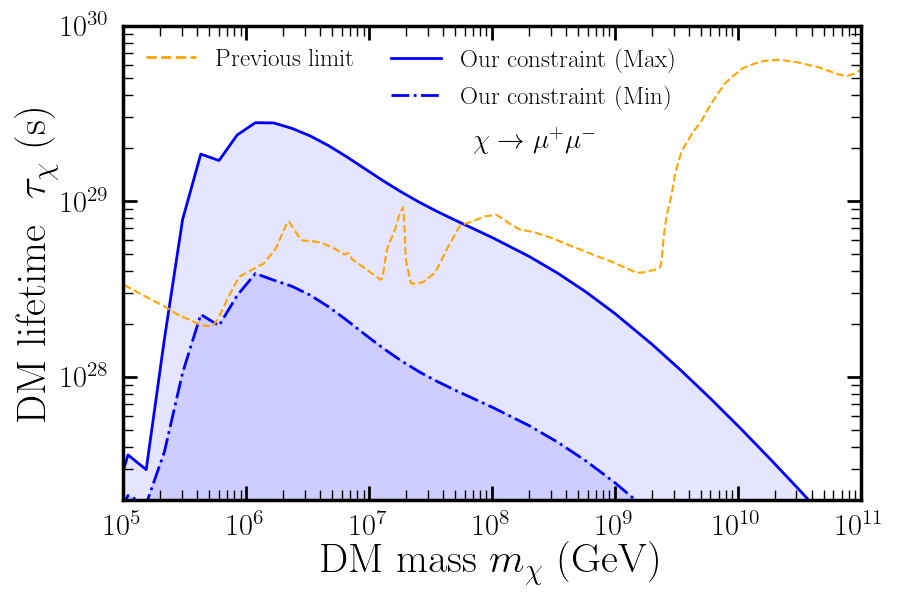}}~~
	\subfloat[\label{sf:limit_decay_nu_e}]{\includegraphics[angle=0.0,width=0.31\textwidth]{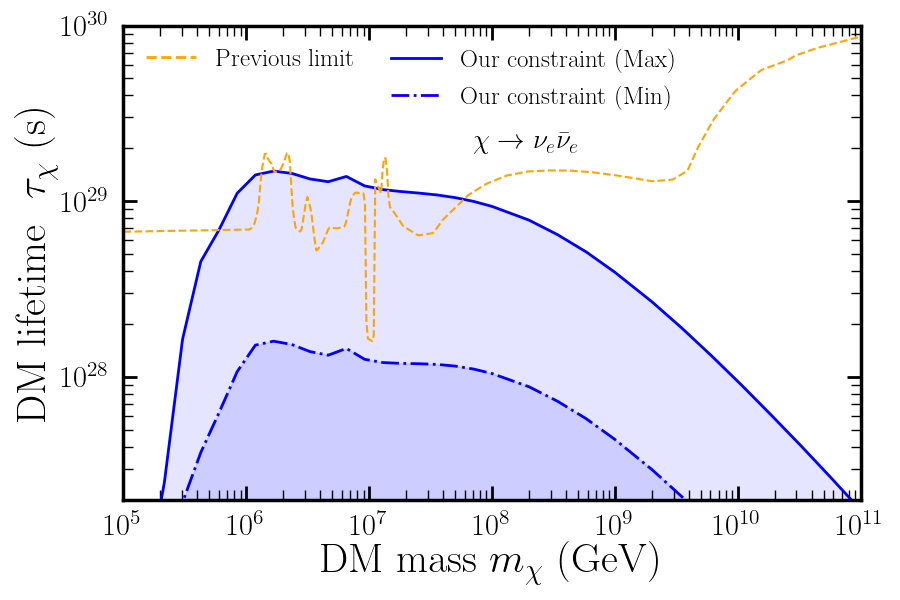}}~~
    \subfloat[\label{sf:limit_decay_nu_mu}]{\includegraphics[angle=0.0,width=0.31\textwidth]{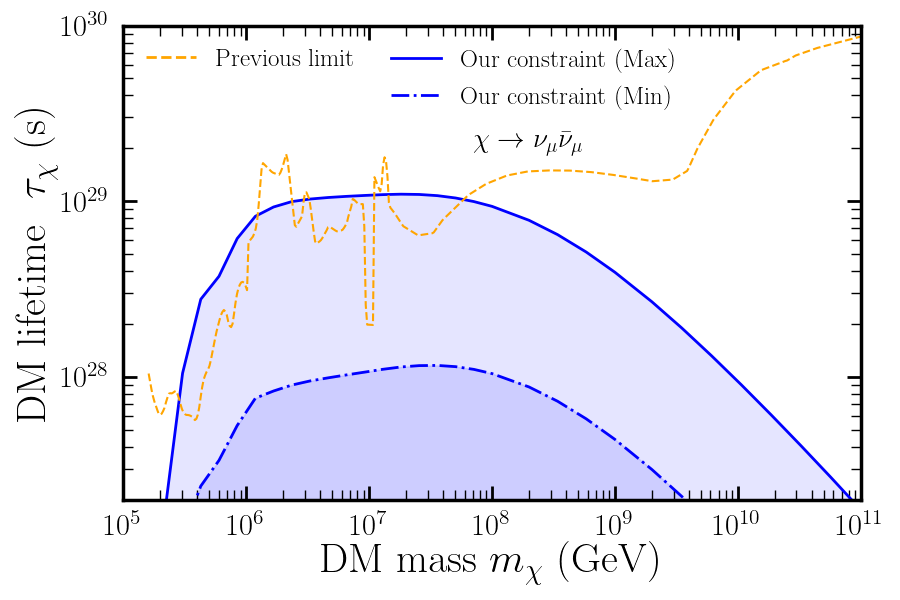}} ~~\\
    \subfloat[\label{sf:limit_decay_nu_tau}]{\includegraphics[angle=0.0,width=0.31\textwidth]{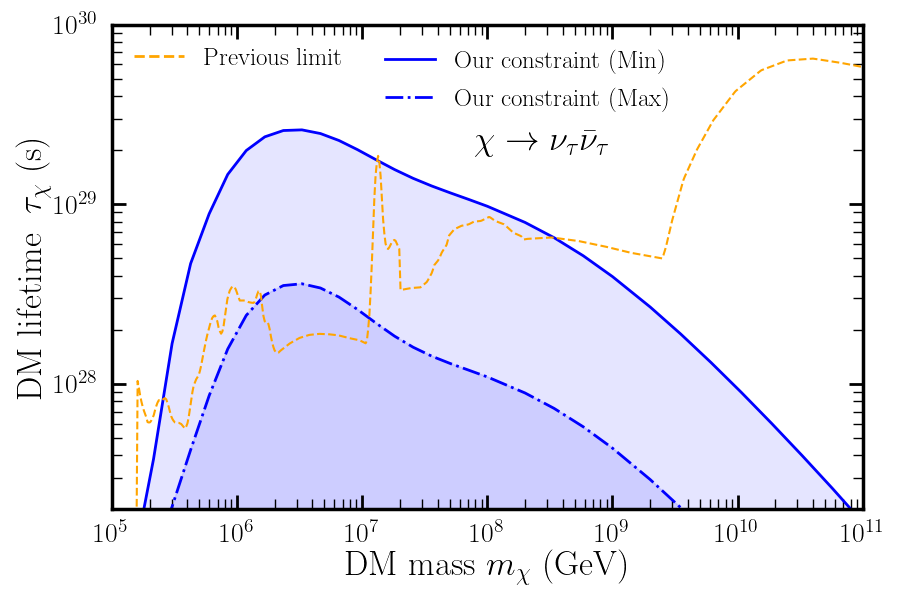}}~~
	\subfloat[\label{sf:limit_decay_H}]{\includegraphics[angle=0.0,width=0.31\textwidth]{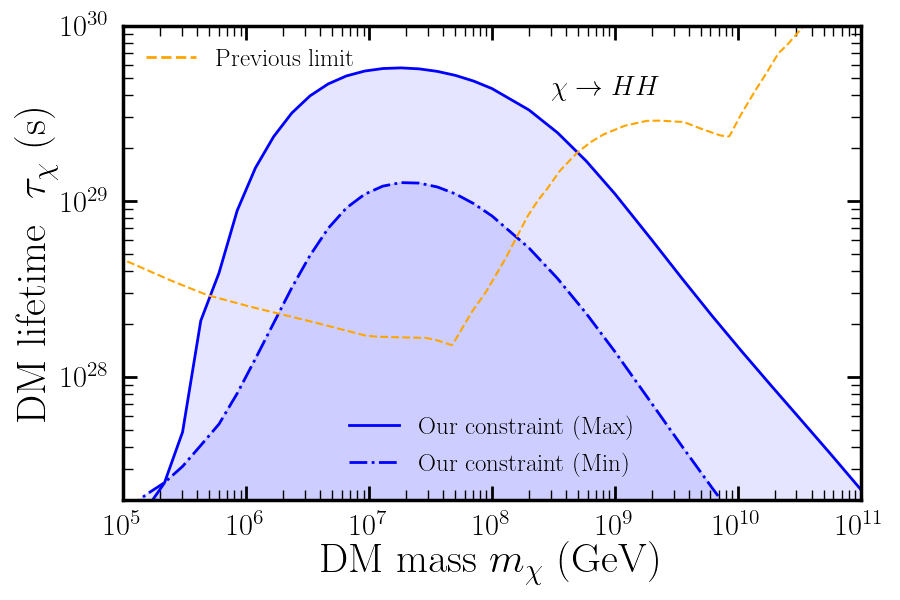}}~~
    \subfloat[\label{sf:limit_decay_Z}]{\includegraphics[angle=0.0,width=0.31\textwidth]{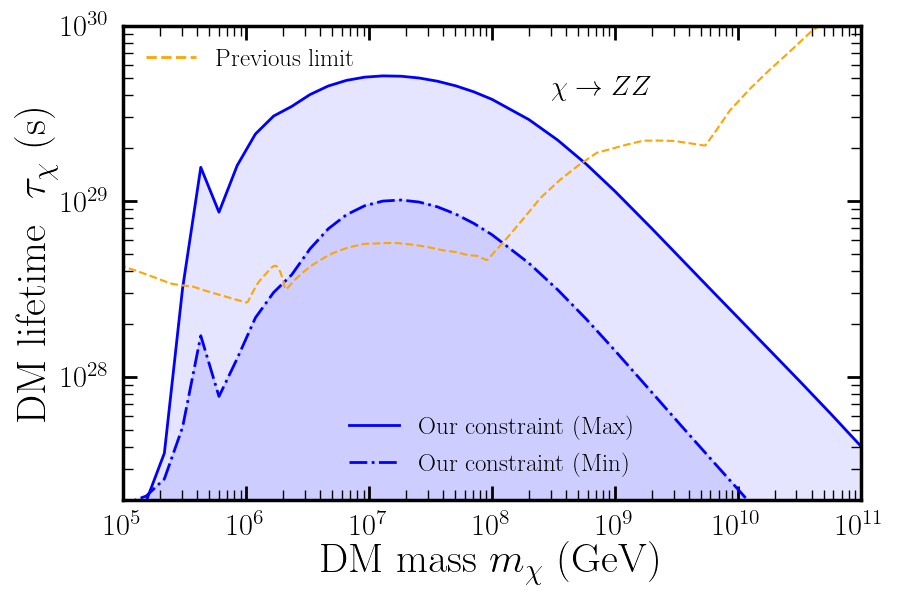}}~~ \\
    \subfloat[\label{sf:limit_decay_W}]{\includegraphics[angle=0.0,width=0.31\textwidth]{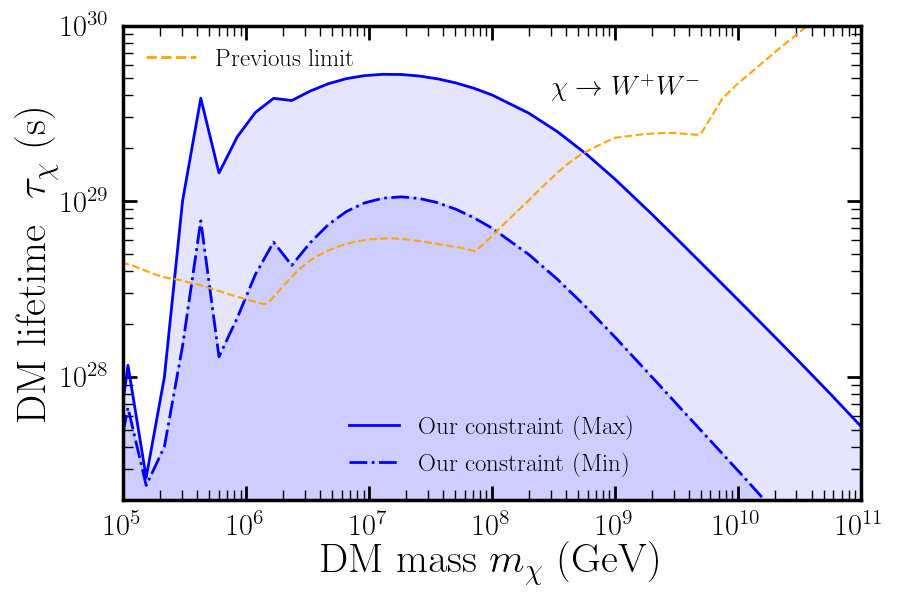}}~~
    \subfloat[\label{sf:limit_decay_gamma}]{\includegraphics[angle=0.0,width=0.31\textwidth]{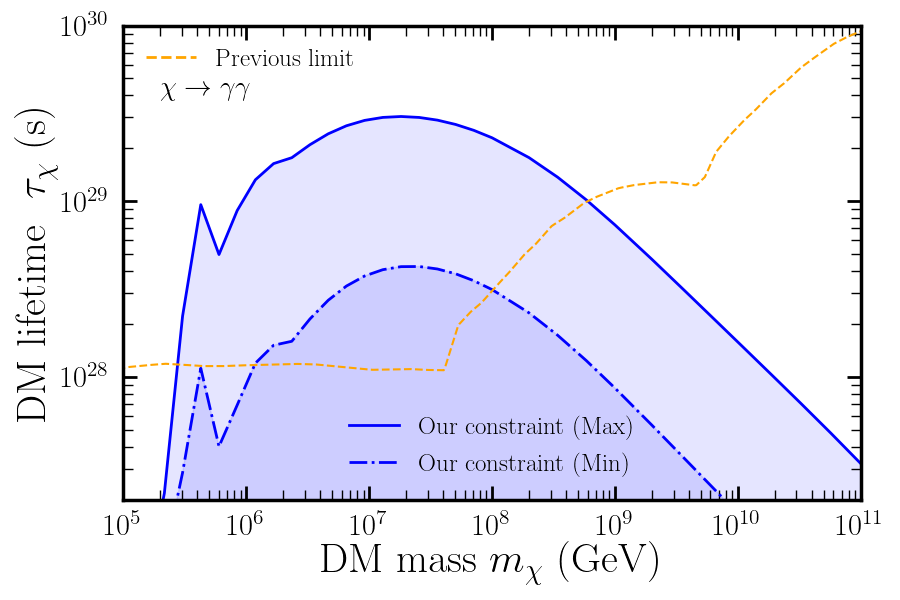}}~~
    \subfloat[\label{sf:limit_decay_g}]{\includegraphics[angle=0.0,width=0.31\textwidth]{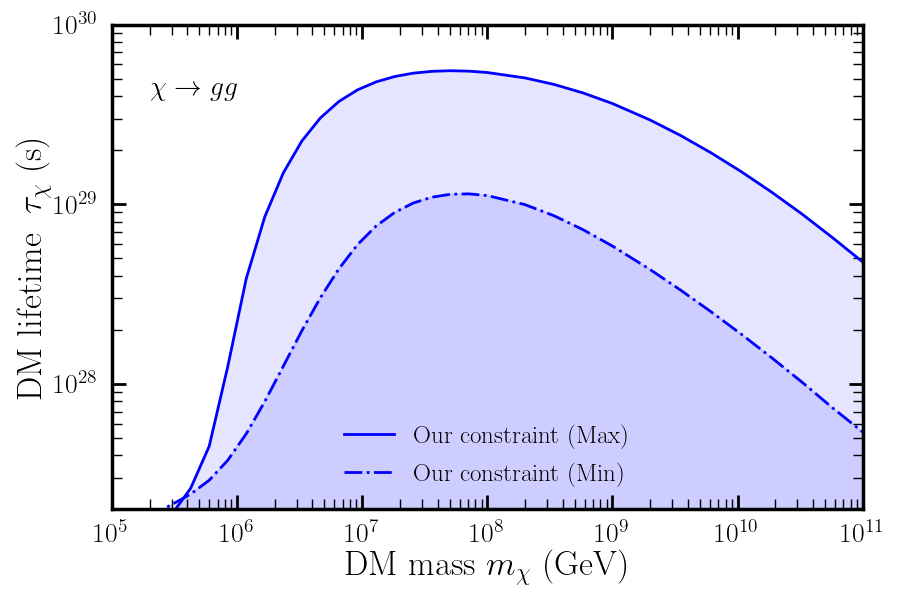}}
    \caption{Limits on decaying DM for other considered final states.}
	\label{fig:limit_decay_other}
\end{center}	
\end{figure*}

%%%%%%%%%%%%%%%%%%%%%
In Fig.\,\ref{fig:limit_decay_other}, we show the limits on decaying DM for other considered two body SM final states. Overall, we find stronger bounds compared to existing results for the majority of the two-body final states analyzed. 

%%%%%%%%%%%%%%%%%%%%%
\section{Annihilating DM: other final states }
\label{app:annihilationDM_other}
\begin{figure*}[h!]
\begin{center}
	\subfloat[\label{sf:limit_annihilation-boost_u}]{\includegraphics[angle=0.0,width=0.31\textwidth]{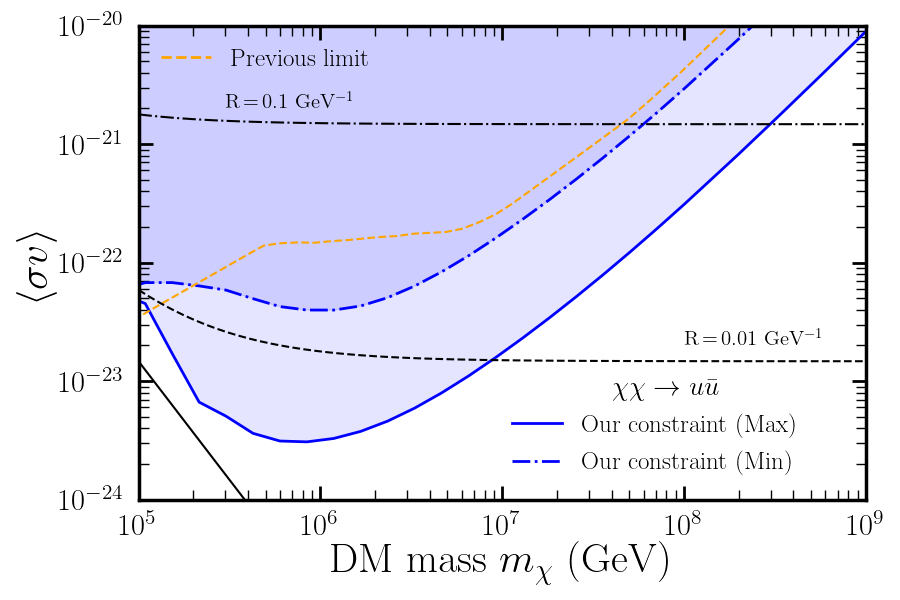}}
	\subfloat[\label{sf:limit_annihilation-boost_d}]{\includegraphics[angle=0.0,width=0.31\textwidth]{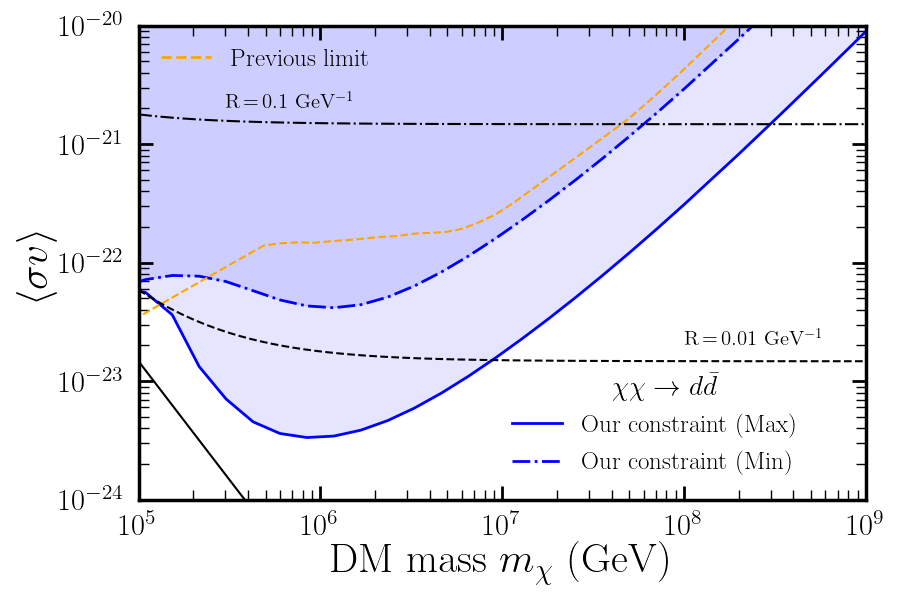}}
    \subfloat[\label{sf:limit_annihilation-boost_c}]{\includegraphics[angle=0.0,width=0.31\textwidth]{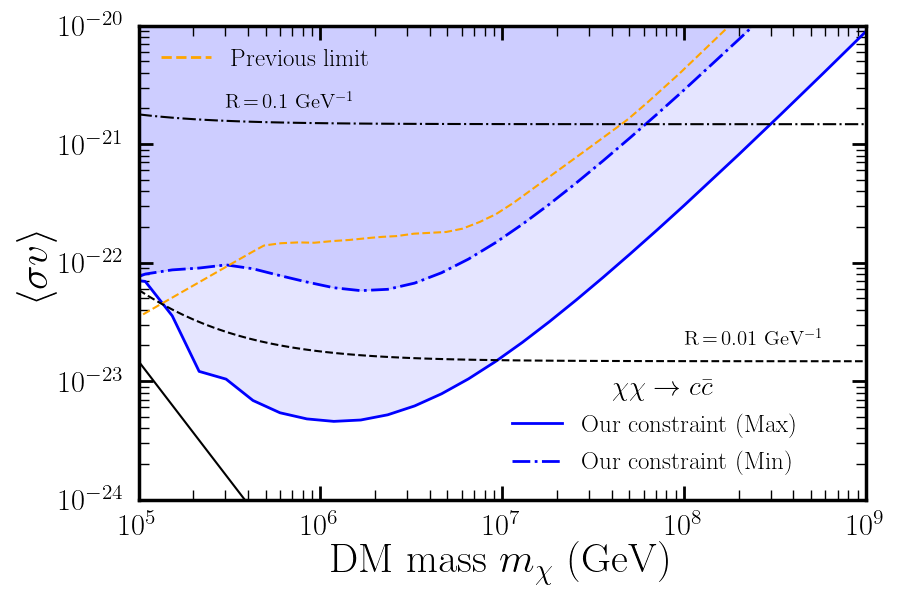}}\\
	\subfloat[\label{sf:limit_annihilation-boost_s}]{\includegraphics[angle=0.0,width=0.31\textwidth]{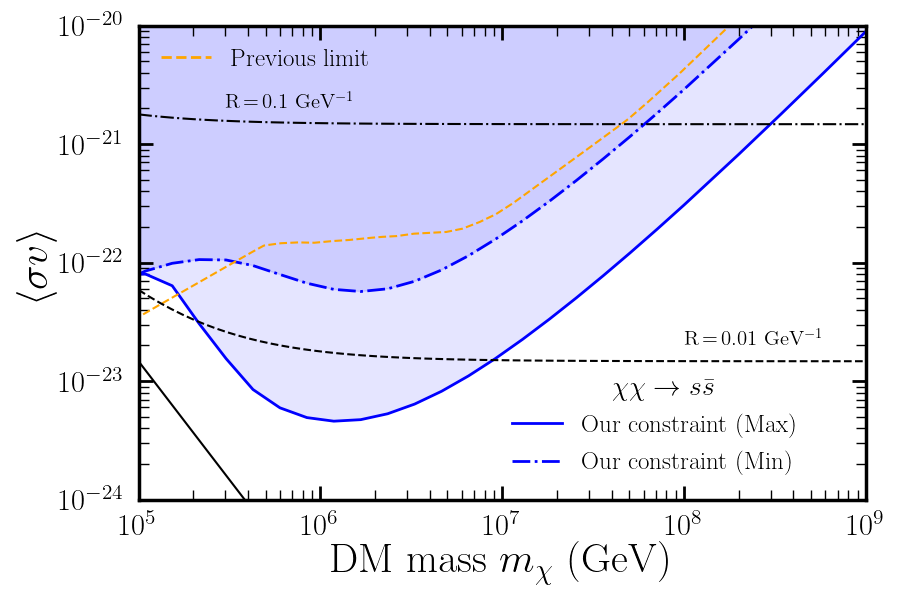}}
	\subfloat[\label{sf:limit_annihilation-boost_t}]{\includegraphics[angle=0.0,width=0.31\textwidth]{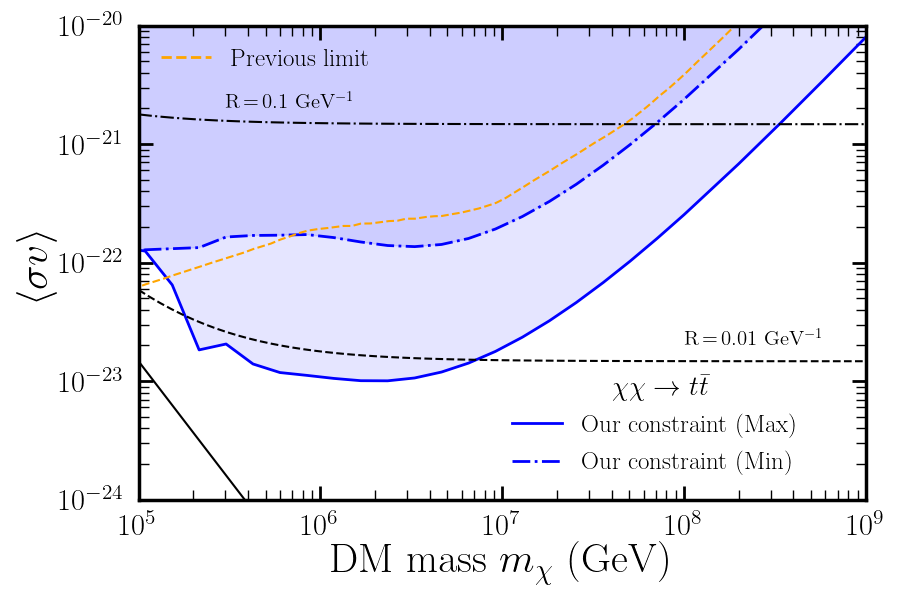}}
    \subfloat[\label{sf:limit_annihilation-boost_e}]{\includegraphics[angle=0.0,width=0.31\textwidth]{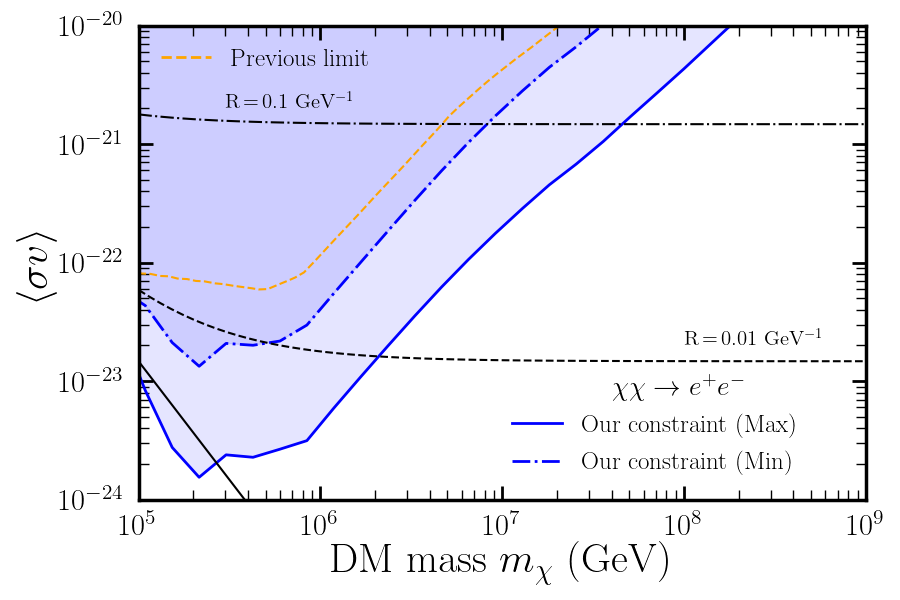}}\\  
    \subfloat[\label{sf:limit_annihilation-boost_mu}]{\includegraphics[angle=0.0,width=0.31\textwidth]{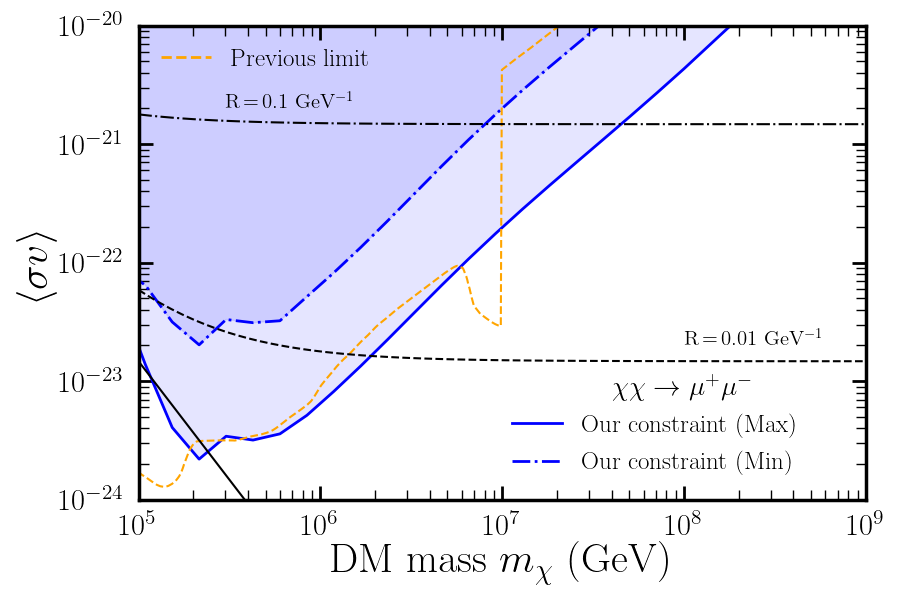}}
	\subfloat[\label{sf:limit_annihilation-boost_nu_e}]{\includegraphics[angle=0.0,width=0.31\textwidth]{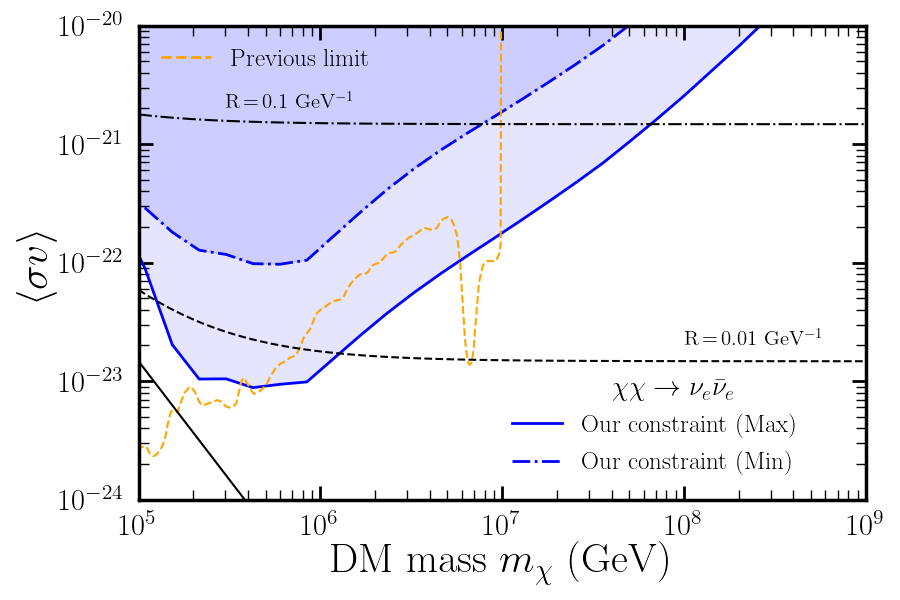}}
    \subfloat[\label{sf:limit_annihilation-boost_nu_mu}]{\includegraphics[angle=0.0,width=0.31\textwidth]{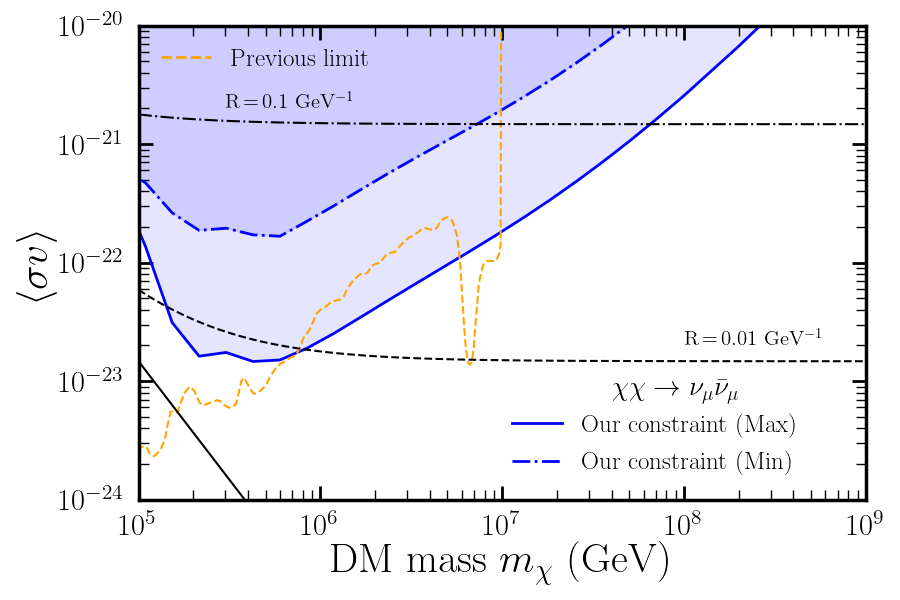}}\\
    \subfloat[\label{sf:limit_annihilation-boost_nu_tau}]{\includegraphics[angle=0.0,width=0.31\textwidth]{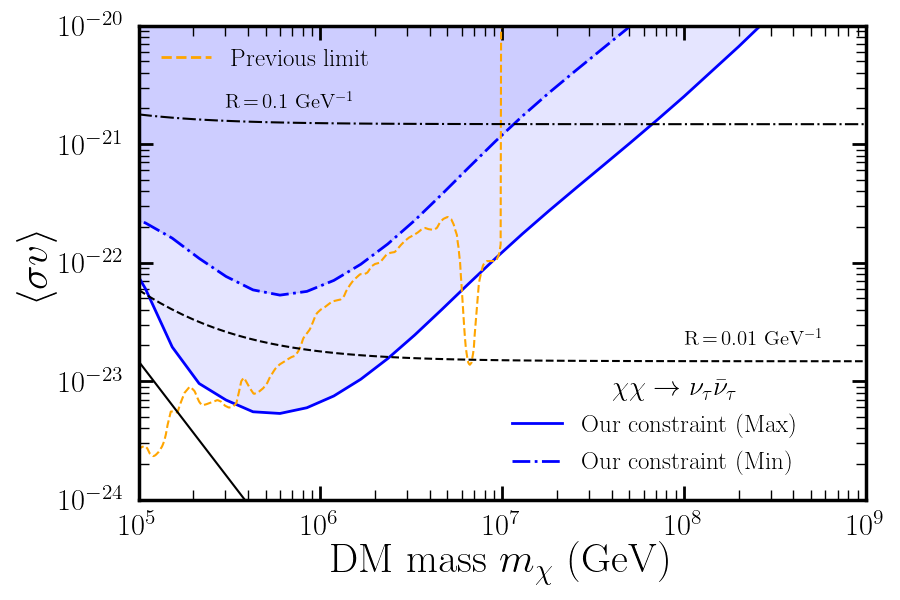}}
	\subfloat[\label{sf:limit_annihilation-boost_H}]{\includegraphics[angle=0.0,width=0.31\textwidth]{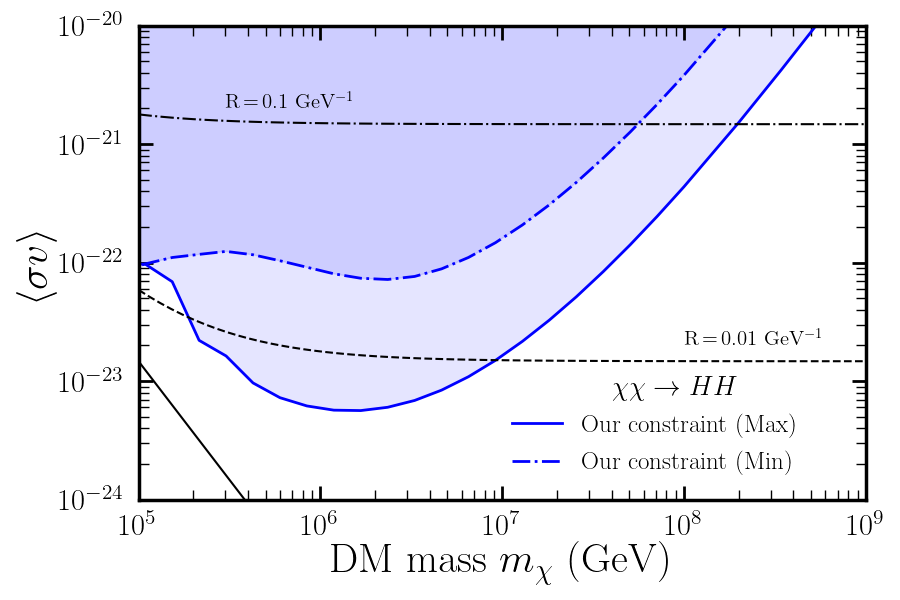}}
    \subfloat[\label{sf:limit_annihilation-boost_Z}]{\includegraphics[angle=0.0,width=0.31\textwidth]{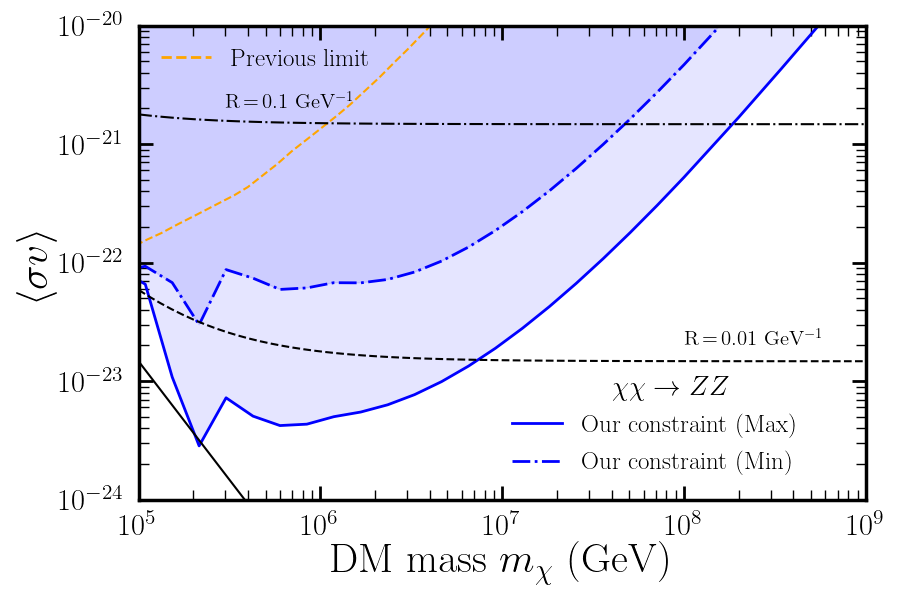}}\\
    \subfloat[\label{sf:limit_annihilation-boost_W}]{\includegraphics[angle=0.0,width=0.31\textwidth]{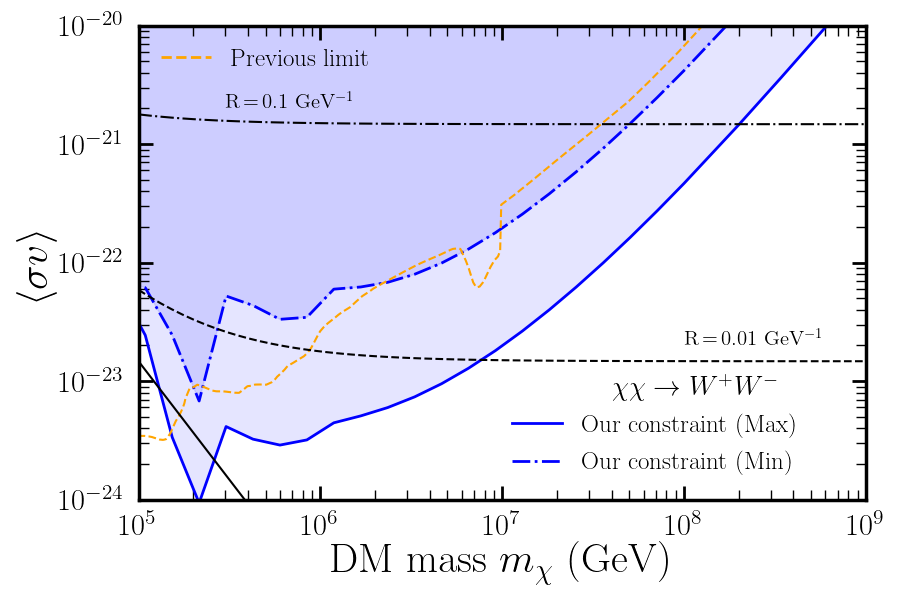}}
    \subfloat[\label{sf:limit_annihilation-boost_gamma}]{\includegraphics[angle=0.0,width=0.31\textwidth]{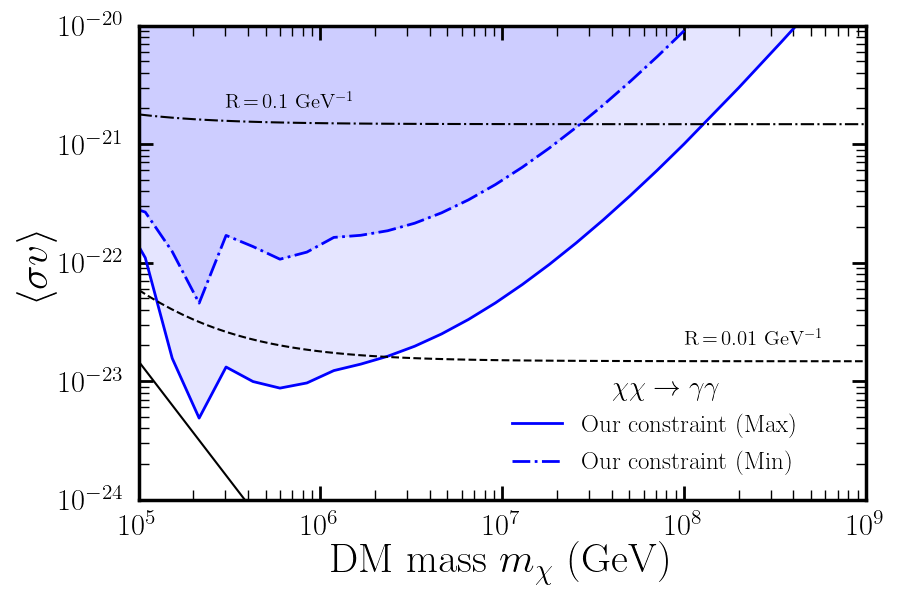}}
    \subfloat[\label{sf:limit_annihilation-boost_g}]{\includegraphics[angle=0.0,width=0.31\textwidth]{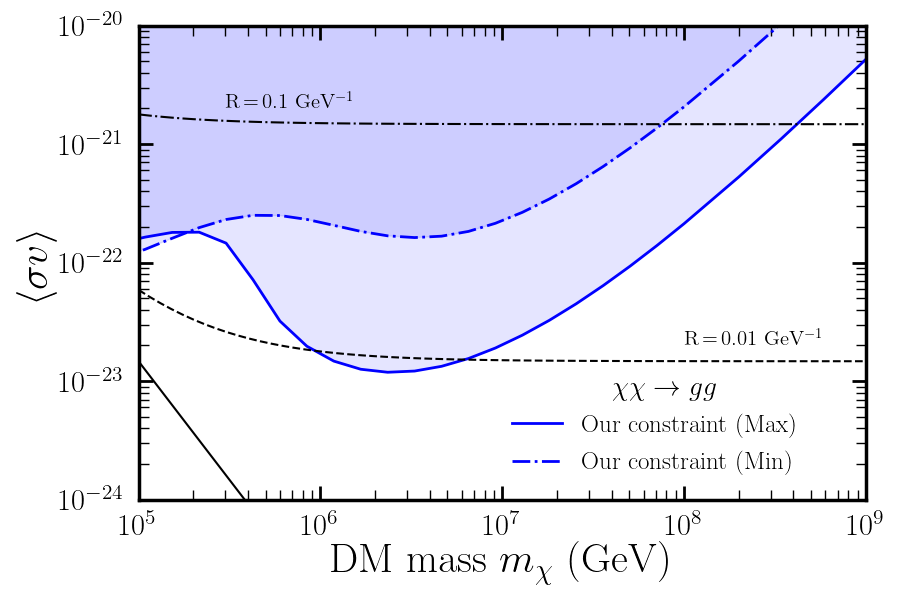}}    
    \caption{Limits on annihilating DM for other considered final states.}
	\label{fig:limit_annihilation_other}
\end{center}	
\end{figure*}

In Fig.\,\ref{fig:limit_annihilation_other}, we present the limits on annihilating DM for other two-body SM final states considered in this work. The limits are computed including boost factor both for galactic and EG case. While the Galactic boost factor is $\sim 1$, the EG boost, though dependent on redshift $z$, can be as large as $\sim 10^5$. Nevertheless, since the EG contribution is subdominant for annihilating DM, this boost does not have a numerically significant impact on our limits. We also show the unitarity bounds for both point-like and composite DM. For previously reported limits, all final states are not always available in the literature. When a limit is available for any of the light quarks, we assume the same limit applies to all light quarks. A similar assumption is made for neutrinos. We do not show previous bounds for the $HH$ channel, as they are not available in the literature for the considered mass range. In general, we obtain stronger limits than those previously reported for most of the two-body final states considered here.  
%%%%%%%%%%%%%%%%%%%%%

\acknowledgments
We thank Marco Cirelli, Dylan M. H. Leung and Kenny C. Y. Ng for useful correspondences. The authors especially thank Abhishek Dubey and Akash Kumar Saha for discussions and coordination of the submission. We thank Deep Jyoti Das, Abhishek Dubey and Akash Kumar Saha  for pointing out an typo in an ealier version of the manuscipt. R.L. acknowledges financial support from the institute start-up funds and ISRO-IISc STC for the grant no. ISTC/PHY/RL/499. The work of T.N.M is supported by the Australian Research Council through the ARC Centre of Excellence for Dark Matter Particle Physics.
\vspace{0.15cm}

\noindent
\textit{Note added:} While our work was in preparation, we became aware of a related study by Dubey et al.\,\cite{Dubey:2025ouh} and Rocamora et al.\,\cite{Rocamora:2025ddt}. Dubey et al.\,\cite{Dubey:2025ouh} utilized the KM2A data set in their analysis, whereas we utilized the combined KM2A and WCDA data sets and adopted a different background model. For the same set of relevant parameters, we find good agreement between our results and those of Rocamora et al.\,\cite{Rocamora:2025ddt}.

\bibliographystyle{JHEP}
\bibliography{ref_LHAASO}

\end{document}